\documentclass[11pt]{article}
\usepackage{cite}
\usepackage{scalerel}
\usepackage{shuffle}
\usepackage{amsmath,amsfonts,amssymb}
\usepackage{latexsym,epsfig}
\usepackage{dsfont}
\usepackage{hyperref}
\usepackage{extarrows}
\usepackage{comment} 
\usepackage{tikz-cd}

\def\hybrid{
        \topmargin -20pt
        \oddsidemargin 0pt
        \headheight 0pt \headsep 0pt
        \textwidth 6.25in 
        \textheight 9.5in 
        \marginparwidth .875in
        \parskip 5pt plus 1pt \jot = 1.5ex}

\hybrid

 \csname
@addtoreset\endcsname{equation}{section}

\def\cO{{\cal O}}
\def\cA{{\cal A}}
\def\cE{{\cal E}}
\def\cM{{\cal M}}
\def\cN{{\cal N}}

\def\cV{{\cal V}}

\def\cK{{\cal K}}

\def\cT{{\cal T}}

\def\del{\partial}
\def\l{\langle}
\def\r{\rangle}

\def\Tr{{\rm Tr}}
\def\B{\square}

\def\e{\epsilon}

\allowdisplaybreaks

\def\bpm{\begin{pmatrix}}
\def\epm{\end{pmatrix}}

\begin{document}

\begin{titlepage}
\rightline{}
\rightline{December 2022}
\rightline{HU-EP-22/41-RTG}  
\begin{center}
\vskip 1.5cm
{\Large \bf{Gauge invariant  double copy of Yang-Mills theory:\\ [1ex]
 the quartic theory}}
\vskip 1.7cm

{\large\bf {Roberto Bonezzi, Christoph Chiaffrino, Felipe D\'iaz-Jaramillo \\ [1.5ex] 
and 
Olaf Hohm}}
\vskip 1.6cm

{\it  Institute for Physics, Humboldt University Berlin,\\
 Zum Gro\ss en Windkanal 6, D-12489 Berlin, Germany}\\[1.5ex] 
 ohohm@physik.hu-berlin.de, 
roberto.bonezzi@physik.hu-berlin.de, 
felipe.diaz-jaramillo@hu-berlin.de, chiaffrc@hu-berlin.de
\vskip .1cm

\vskip .2cm

\end{center}

\bigskip\bigskip
\begin{center} 
\textbf{Abstract}

\end{center} 
\begin{quote}

We give  an explicit gauge invariant, off-shell  and local double copy 
construction of gravity from Yang-Mills theory to quartic order. 
To this end we use the framework of homotopy algebras, and we 
identify a rich new algebraic structure associated to color-stripped 
Yang-Mills theory. This algebra, which is a generalization of a Batalin-Vilkovisky algebra,  
is  the underlying structure necessary 
for double copy. 
We give a self-contained introduction into these algebras 
by illustrating them  for Chern-Simons theory in three dimensions. We then construct   
 $N=0$ supergravity in the form of double field theory in terms of the algebraic  Yang-Mills 
building blocks to quartic order in  interactions. 
As applications of the same universal formula, we re-derive the 4-graviton scattering amplitude and 
compute a chiral form of 
the Courant  algebroid gauge structure of double field theory.

\end{quote} 
\vfill
\setcounter{footnote}{0}
\end{titlepage}

\tableofcontents

\vspace{5mm}
\section{Introduction}

Double copy is a powerful technique to compute  gravity scattering amplitudes from 
gauge theory amplitudes. Originally discovered in string theory \cite{Kawai:1985xq}, the first double copy  construction in field theory, proposed by Bern, Carrasco and Johansson (BCJ) \cite{Bern:2008qj}, 
relates Yang-Mills theory to Einstein-Hilbert gravity coupled to 
an antisymmetric tensor ($B$-field) and a scalar (dilaton). This gravity theory is 
commonly referred to as $N=0$ supergravity and, in view of double copy, is most 
efficiently formulated as a 
double field theory \cite{Siegel:1993th,Hull:2009mi,Hull:2009zb,Hohm:2010jy,Hohm:2010pp,Hohm:2011dz,Aldazabal:2013sca,Berman:2013eva,Hohm:2013bwa}.

The double copy program relies on a feature of gauge theory known as 
`color-kinematics duality'\cite{Bern:2008qj,Bern:2019prr}. This refers to the  at first `experimental' observation that 
the kinematic numerators of Yang-Mills theory can be brought to a form 
where they obey the same relations as the color factors  built from structure constants. 
Since for the color factors these relations follow from the Jacobi
identities of the color Lie algebra, this observation suggests that
there is a hidden `kinematic' Lie algebra. 
Despite partial progress \cite{Monteiro:2011pc,Chen:2019ywi,Reiterer:2019dys,Brandhuber:2021bsf,Ben-Shahar:2021zww}, this kinematic Lie algebra has remained elusive. 
Nevertheless, color-kinematics duality has been proved for 
tree-level amplitudes by various indirect methods \cite{Bjerrum-Bohr:2010pnr,Mafra:2014oia,Mafra:2015vca,Lee:2015upy}. Moreover, double copy has been tested and applied with 
great success for loop level amplitudes \cite{Bern:2012uf,Bern:2017yxu,Bern:2018jmv} and, more recently, for classical solutions \cite{Monteiro:2014cda,Luna:2015paa,Luna:2016hge,Luna:2018dpt,Arkani-Hamed:2019ymq,Monteiro:2021ztt,Godazgar:2022gfw} including the two-body 
problem relevant for black hole inspiral \cite{Goldberger:2016iau,Shen:2018ebu,Bern:2019crd,Bern:2020buy,Shi:2021qsb}. 
Double copy thus promises a profound new outlook on classical and quantum gravity, 
but  we are still lacking the kind of  first-principle 
understanding  that would be necessary in order to delineate  the exact scope of 
double copy.

It is therefore highly desirable to have an off-shell derivation of  color-kinematics duality and double copy 
starting from a fundamental formulation of the gauge theory, such as  its Lagrangian (see \cite{Bern:2010yg,Borsten:2020xbt,Diaz-Jaramillo:2021wtl} for Lagrangian double copy constructions). 
In this paper we present an explicit double copy construction, up to and including quartic couplings,  for 
Yang-Mills theory in $D$-dimensional Minkowski spacetime and recover double field theory (DFT) to this order. 
Importantly, our double copy procedure is manifestly 
off-shell, gauge invariant and local. 
In particular locality is  important  in order to eventually prove   loop-level color-kinematics duality, while gauge invariance is 
desirable for treating classical solutions in a manner that avoids arbitrary gauge choices.

To this end we build on our previous work using the homotopy algebra formulation of 
general gauge field theories \cite{Zwiebach:1992ie,Hohm:2017pnh} in order to double copy Yang-Mills theory 
to cubic order \cite{Bonezzi:2022yuh}, which here we  generalize to quartic order.  
The transition to quartic order is indeed a critical test of any double copy construction, as 
for instance the algebra of gauge transformations and its Jacobi identity 
become  first  visible to this  order. 
Following previous important work by Reiterer \cite{Reiterer:2019dys}, we identify a vast hidden 
algebraic structure associated to the kinematics of Yang-Mills theory 
and use it to construct gauge invariant gravity (in the form of double field theory) 
to quartic order.  
This kinematic algebra is a homotopy generalization  of a Batalin-Vilkovisky (BV) algebra, 
which was proposed by Reiterer as the algebra explaining color-kinematics duality 
for Yang-Mills theory (at least in four dimensions in Euclidean signature).\footnote{Specifically, Reiterer employs 
a first-order  formulation  of Yang-Mills theory that requires self-dual two-forms and is hence specific to four dimensions, 
see also Costello \cite{costellorenormalization}.}
We have every reason to believe that eventually this double copy construction of gravity  from Yang-Mills theory 
will be extendable  to all orders. 
Accomplishing  this program will   amount to a complete first-principle understanding of double copy.

The algebraic 
structures to be discussed in this paper  must  appear rather arcane to the general theoretical physicist  (they certainly 
did so to the present authors not too long ago), but we will give a completely self-contained introduction. 
In particular, one can illustrate all essential ingredients in the much simpler context of three-dimensional 
Chern-Simons theory, for which at least part of the kinematic Lie algebra was recently identified 
by Ben-Shahar and Johansson in \cite{Ben-Shahar:2021zww}. Remarkably, Chern-Simons theory 
shows precisely the same structure as  Yang-Mills theory, with the one exception that for the latter the algebraic structures 
are generically `up to homotopy',  a notion that we will explain shortly.

In the remainder of the introduction we briefly sketch the main technical  ingredients needed for our 
double copy construction, and  we describe how to push it  beyond cubic order.
In the framework of homotopy algebras one encodes a Lagrangian field theory in terms 
of a cyclic $L_{\infty}$ algebra (also called  strongly homotopy Lie algebras \cite{Lada:1992wc}), for which the action for fields $\psi$ reads 
\cite{Hohm:2017pnh} 
 \begin{equation}\label{firstLinftyAction}
  S  = \frac{1}{2} \big\langle \psi , B_1(\psi)\big\rangle  + \frac{1}{3!} \big\langle \psi, B_2(\psi,\psi)\big\rangle 
  +\frac{1}{4!} \big\langle \psi, B_3(\psi,\psi,\psi)\big \rangle +\cdots\;. 
 \end{equation} 
 Here $B_1$, $B_2$, $B_3$, etc., are multi-linear maps of fields (and gauge parameters and other 
 data defining a field theory), while $\langle \, \,,\, \rangle $ denotes an inner product. 
The $L_{\infty}$ maps are subject to quadratic generalized Jacobi identities, and the inner product 
obeys suitable cyclicity conditions, which together ensure gauge invariance 
and general consistency conditions of the field theory. 
It must be emphasized that, for concrete theories,  the $L_{\infty}$ maps $B_1$, $B_2$, etc., 
are just local expressions of fields  so that (\ref{firstLinftyAction}) reproduces 
the familiar actions of field theory. The point is simply that the above  provides an algebraic formulation 
of the consistency conditions of gauge field theories in terms of $L_{\infty}$ algebras \cite{Hohm:2017pnh}. 

Given this algebraic perspective  one can give a perfectly meaningful interpretation of 
`color-stripping' the action of Yang-Mills theory. By writing out the color indices 
in the $L_{\infty}$ maps of Yang-Mills theory, the latter can be written as 
 \begin{equation}\label{YMactionIntro}
  S_{\rm YM} = \frac{1}{2}\big\langle A_{a}, m_{1}(A^{a})\big\rangle +\frac{1}{6} f_{abc} \big\langle 
  A^a, m_2(A^b, A^c)\big\rangle 
  +\frac{1}{12} f_{abe} f_{cd}{}^{e} \big\langle A^a, m_3(A^b, A^c, A^d)\big\rangle\;, 
 \end{equation}  
where $f_{abc}$ are the structure constants of the color Lie algebra. 
This gives rise to linear, bilinear and trilinear maps $m_{1}$, $m_2$ and $m_3$, respectively, 
which define an algebra of their own without any color structure. 
We have thus split the vector space of fields into a tensor product ${\cal K}\otimes \mathfrak{g}$, 
where $\mathfrak{g}$ denotes the color Lie algebra, while ${\cal K}$ is the space of `kinematic' 
Yang-Mills structures. 
These maps satisfy relations like $m_{1}^2=0$, which together with the Jacobi identity for $f_{abc}$ 
imply gauge invariance of the action. More precisely, $m_{1}$, $m_2$ and $m_3$ define on ${\cal K}$ 
a graded commutative algebra up to homotopy, called  $C_{\infty}$ algebra for short \cite{Zeitlin:2008cc}. 
This means in particular that $m_2$ defines a graded commutative product that is associative 
up to corrections governed by $m_{1}$ and $m_3$.  We note that three-dimensional Chern-Simons theory 
takes the same form (\ref{YMactionIntro}) except that in this case $m_3$ vanishes, while $m_{1}$, which involves 
$\square$ for Yang-Mills theory, reduces to the de Rham differential.  Consequently, 
the $C_{\infty}$ algebra of Chern-Simons theory is just  the associative algebra of differential forms.

The Yang-Mills action  we will use 
is  of the standard textbook form, except that it features  one auxiliary scalar in order 
to isolate a kinetic term involving the d'Alembert operator $\square$ in a gauge invariant manner. 
This has the important consequence that there is a second nilpotent operator $b$, of opposite 
degree to $m_{1}$, that acts in  a purely algebraic manner and satisfies 
 \begin{equation}\label{QBox} 
  m_{1} \, b + b\,  m_{1} = \square\;. 
 \end{equation}
This second `differential' does not, however, act as a derivation on the product $m_2$. 
Rather, the failure of $b$ to do so defines a new structure. Setting 
 \begin{equation} 
  b_2 = bm_2 - m_2 b\,, 
 \end{equation} 
with a notation to be made precise below, one obtains a bracket $b_2$
that fails to obey the Jacobi identity, and hence  to define a Lie algebra, by certain controlled maps. 
A new source of failure originates  from (\ref{QBox}) and denotes terms 
of the structural form $A\partial^{\mu} B \partial_{\mu}C$. 
An algebraic structure encoding these so-called $\square$--failures 
was proposed by Reiterer \cite{Reiterer:2019dys}, and following his terminology  we 
refer to it as a BV$_{\infty}^{\square}$ algebra. This indeed appears to be the structure extending color-kinematics duality and enabling double copy beyond scattering amplitudes. In particular, we will show that a compatibility condition between the bracket $b_2$ and the product $m_2$, which is part of the ${\rm BV}_\infty^\B$ axioms, reduces to the iconic relation ${n_s+n_t+n_u=0}$ for the kinematic numerators of the 4-point Yang-Mills amplitude.

A BV$_{\infty}^{\square}$ algebra is present purely on the kinematic vector space ${\cal K}$ 
of Yang-Mills theory, as 
we will prove by computing the explicit maps up to and including trilinear maps. (Unfortunately, we are not aware of a compact
definition of BV$_\infty^\B$ algebras with maps carrying an arbitrary
number of inputs). 
Based on this we can double copy by introducing a second copy $\bar{\cal K}$, 
whose corresponding maps are denoted by a bar,  and 
taking the tensor product ${\cal K}\otimes \bar{\cal K}$. This space consists of functions 
of a doubled set of coordinates, say $x$ associated to ${\cal K}$ and $\bar{x}$ associated 
to $\bar{\cal K}$. Restricting to the subspace that is annihilated by $b^- := \frac12(b\otimes 1 - 1\otimes \bar{b})$, 
and restricting to functions that 
are `strongly constrained' in the sense of DFT, so that $\square=\bar{\square}$ acting on any functions 
and products of functions, 
one can determine  an $L_{\infty}$ structure, and hence a consistent classical field theory, in the form of DFT.   
In this the $\square$--failures  on the Yang-Mills side translate 
to `failures by $\square-\bar{\square}$' on the gravity side, but here they are eliminated by 
the  `section constraints' of DFT, 
giving rise to a genuine $L_{\infty}$  algebra. 
Specifically, the $L_{\infty}$  maps  $B_1$, $B_2$, $B_3$, etc., defining the  DFT 
action in the form (\ref{firstLinftyAction}),  are defined from the Yang-Mills ingredients as follows: 
The differential is given by $B_1= m_{1}+\bar{m}_{1}$, while the 2-bracket can be written as 
 \begin{equation}
  B_2 = \frac{1}{4} \left(m_2\otimes \bar{b}_2 - b_2\otimes \bar{m}_2\right)\;. 
 \end{equation} 
Note that for the special case that $m_2$ defines a strictly commutative associative algebra 
and that $b_2$ defines a Lie algebra, each term here takes the form of a familiar tensor product 
of a commutative times a Lie algebra, giving a new Lie algebra. Since $m_2$ and  $b_2$  are not 
strict, the above $B_2$ does not define a genuine Lie algebra, but it defines an $L_{\infty}$ algebra whose 
$B_3$ can be 
expressed in terms  of the Yang-Mills ingredients, giving a result of the schematic form 
$B_3 \propto b_3\otimes \bar{m}_2\bar{m}_2 + m_2m_2\otimes \bar{b}_3+\cdots$. 
We give the explicit algebraic formula  for $B_3$ in eq. \eqref{niceB3} below, which is one 
of the core technical results of this paper. 
This formula  for $B_3$ encodes not only the quartic interactions but 
all data relevant for the quartic theory, such as  
the 3-bracket of the higher gauge algebra of DFT.   
We test and apply this formula, first, by computing 
the 3-bracket of the Courant-type gauge algebra of DFT in a chiral basis and, second, by re-deriving 
the 4-graviton scattering amplitude in terms of squares of Yang-Mills amplitudes. 

The rest of this paper is organized as follows. In sec.~2 we take the opportunity to 
introduce the `strict version' of these algebraic structures by reviewing Chern-Simons theory 
and its recently identified kinematic Lie algebra \cite{Ben-Shahar:2021zww}. We then turn in sec.~3 to genuine Yang-Mills theory 
and identify the BV$_{\infty}^{\square}$ algebra on its kinematic vector space ${\cal K}$, 
displaying and proving its defining relations up to and including trilinear maps. 
These results are used in sec.~4 to double copy Yang-Mills theory by re-deriving  $B_2$ and 
computing the new $B_3$. Furthermore, we test our algebraic formula for $B_3$ by 
computing the 4-graviton amplitude and the 3-bracket of the gauge algebra. 
In sec.~5 we close with brief conclusions and an outlook, while in appendix A we collect all maps 
of the BV$_{\infty}^{\square}$ algebra, and in appendix B we give a self-contained summary of 
BV$_{\infty}$ algebras \cite{Galvez-Carrillo:2009kic} without $\square$--failures.

\section{Chern-Simons theory as a BV algebra}

In this section we review three-dimensional Chern-Simons theory  and its kinematic Lie algebra, 
which was recently uncovered by Ben-Shahar and Johansson \cite{Ben-Shahar:2021zww}, as a way of introducing the 
strict versions of the algebraic structures to be employed below for Yang-Mills theory. 
In this we only assume familiarity with differential forms. 

\subsection{Chern-Simons theory} 
Differential forms form a vector space that in three dimensions is given by 
$\Omega^{\bullet}=\bigoplus_{p=0}^3 \Omega^p$, 
where $\Omega^p$ is the space of $p$-forms. 
(Here one permits the sum of differential 
forms of different degrees, but usually it is understood that we consider homogeneous elements of fixed degree.) 
One says that $\Omega^{\bullet}$ carries an \textit{integer grading} given by the form degree, 
and   further that it is a \textit{chain complex}:  a sequence of vector spaces connected by a map $d$ 
(the differential) acting as 
\begin{equation}\label{3DdeRham} 
\begin{array}{cccccc}
0&\xlongrightarrow\;\; \; \Omega^0&\xlongrightarrow{d}\;\; \; \Omega^{1}&\xlongrightarrow{d}\;\; \;\Omega^{2}&\xlongrightarrow{d}\;\; \;\Omega^{3}&\xlongrightarrow\;\; \; 0 \;, 
\end{array}  
\end{equation}
where    $d^2=0$. For differential forms, $d$ is the de Rham differential acting in the familiar fashion via $d=dx^{\mu}\partial_{\mu}$, 
e.g., for a one-form $u=u_{\nu}dx^{\nu}$ we have $d u  = \partial_{\mu}u_{\nu}
dx^{\mu}\wedge dx^{\nu}$. 
The chain complex (\ref{3DdeRham}) is known as the de Rham complex (in three dimensions). 

The de Rham differential is a linear map, but importantly the de Rham complex also carries a non-linear 
algebraic structure given by the {wedge product} $\wedge$. It is defined in the familiar fashion, e.g.~on one-forms 
$u_{1,2} = u_{1,2\,\mu} dx^{\mu}$ as  $u_1\wedge u_2=u_{1\mu}u_{2\nu} dx^{\mu}\wedge dx^{\nu}$. 
The wedge product is associative, obeys a Leibniz rule with respect to $d$, and is \textit{graded symmetric}. 
In order to display these relations in an abstract form that makes the generalization to homotopy versions below 
more transparent, we now set for arbitrary $u, v \in \Omega^{\bullet} $
 \begin{equation}
 \begin{split}
  m_1(u) := du \;, \qquad\; 
  m_2(u,v) := u \wedge v\;. 
 \end{split} 
 \end{equation} 
Denoting the degree of a (homogeneous) element $u$ by $|u|$, i.e., for $u\in \Omega^p$ we have $|u|=p$, 
the above maps obey $|m_1(u)|=|u|+1$ and $|m_2(u,v)|=|u|+|v|$. We then say  that the 
intrinsic degrees of $m_1, m_2$ are $|m_1|=1$ and $|m_2|=0$, respectively. 
The product is graded symmetric or \textit{graded commutative} in the sense that 
 \begin{equation}\label{gradecomm}
  m_2(u_1,u_2) = (-1)^{u_1 u_2} m_2(u_2,u_1) \;, 
 \end{equation}
where in exponents we use the short-hand notation $(-1)^{u_1u_2}\equiv (-1)^{|u_1||u_2|}$. 
This relation expresses the (anti-)commutativity of the wedge product. 
Similarly, the nilpotency of $d$, the Leibniz rule between $d$ and $\wedge$, 
and the associativity $(u\wedge v)\wedge w=u\wedge (v\wedge w)$ now read 
 \begin{equation}\label{strictCinfty} 
  \begin{split}
   m_1^2 &= 0\;, \\
   m_1(m_2(u,v)) - m_2(m_1(u),v)- (-1)^{u}m_2(u,m_1(v)) &=0 \;, \\
   m_2(m_2(u,v),w) - m_2(u,m_2(v,w)) &= 0\;. 
  \end{split}
 \end{equation} 
In general, a  chain complex with differential $m_1$  equipped with a graded commutative and associative map $m_2$ 
satisfying (\ref{gradecomm}), (\ref{strictCinfty}) is called  a \textit{differential graded commutative algebra} (dgca), 
which is a special case (the \textit{strict} version) of a $C_{\infty}$-algebra.

We next turn to Chern-Simons theory and 
 introduce the Lie algebra $\mathfrak{g}$ of its  `color' gauge group. We denote 
the  structure constants by $f_{ab}{}^{c}$, which obey the Jacobi identity $f_{[ab}{}^{d} f_{c]d}{}^{e}=0$, 
and the generators by $t_a$.  
One then defines a new chain complex $X^{\bullet}=\bigoplus_{i=0}^{3} X^{i}$, where 
 \begin{equation}\label{Xtensor}
  X^i := \Omega^{i}\otimes \mathfrak{g}\;. 
 \end{equation} 
By this  we just mean that the differential forms are promoted to forms taking values in the Lie algebra $\mathfrak{g}$. 
For instance, for a $\mathfrak{g}$-valued one-form we write 
\begin{equation}
  A=  A_{\mu}{}^{a}dx^{\mu}\otimes  t_a  = A^a t_a \in X^1\;, 
 \end{equation} 
with the understanding that $A^a$ is a one-form.  
The differential $m_1=d$ extends to a differential  on 
$X^{\bullet}$ that we also call $d$ and that acts as $dA =d(A^a)\otimes t_a$ 
and of course still  obeys $d^2=0$. 
Similarly, 
the `2-product' $m_2$ of the dgca extends to  a `2-bracket' defined by 
 \begin{equation}\label{tensorbracket}
  [A_1,A_2 ]  :=  m_2(A_1^b, A_2^c) f_{bc}{}^{a} \otimes t_a\;, 
 \end{equation}
which due to the structure constants is now \textit{graded antisymmetric}, 
 \begin{equation}
  [A_1,A_2] = (-1)^{A_1A_2+1} [A_2, A_1]\;. 
 \end{equation}  
Thanks to the dgc structure on $\Omega^{\bullet}$, and the Lie algebra structure of $\mathfrak{g}$, the above 
complex inherits the structure of a \textit{differential graded Lie algebra} (dgLa), 
which is a special case (the \textit{strict} version) of an $L_{\infty}$-algebra. This means that $d$ and $[\,\cdot\,, \,\cdot\,] $ 
obey 
 \begin{equation}\label{strictLinfty} 
  \begin{split}
   d^2 &= 0\;, \\
   d[A_1,A_2] - [dA_1, A_2] - (-1)^{A_1} [A_1,dA_2]  &=0 \;, \\
   [[A_1,A_2],A_3] + (-1)^{A_1(A_2+A_3)} [[A_2,A_3],A_1]+(-1)^{A_3(A_1+A_2)}  [[A_3,A_1],A_2] &= 0\;, 
  \end{split}
 \end{equation} 
the last relation being  the graded Jacobi identity.

Given a dgLa (or in fact an $L_{\infty}$-algebra) one can define a (classical) field theory, which 
has an action provided there is an inner product or pairing $\langle \cdot\, , \cdot\rangle : X^i\otimes X^{3-i} \rightarrow \mathbb{R}$
for $i=0,1,2,3$, 
obeying the `cyclicity conditions'  that 
 \begin{equation}\label{cyclicity} 
  \langle A_1,[A_2, A_3] \rangle \qquad  \text{and} \qquad 
  \langle A_1, dA_2\rangle 
  \end{equation}
are completely graded antisymmetric. 
This implies, in particular,  that if  $A_{1}, A_{2}, A_{3}$ are all of degree one  the tri-linear object 
$\langle [A_1, A_2], A_3\rangle$ is totally symmetric under permutations of  $123$. 
For the above dgLa an inner product exists whenever the Lie algebra $\mathfrak{g}$ carries an 
invariant quadratic form $\kappa_{ab}$, because for $A\in X^i, B\in X^{3-i}$ we can integrate 
the three-form $A\wedge B$ over the 3-manifold underlying the de Rham complex: 
\begin{equation}
  \langle A, B\rangle := \kappa_{ab}  \int A^a\wedge B^b\;. 
 \end{equation}
Cyclicity  follows by discarding total derivatives and using that $\kappa_{ab}$ is invariant, which in turn implies 
that $f_{abc}:=\kappa_{ad}f^{d}{}_{bc}$ is totally antisymmetric.

We can now write the Chern-Simons action for a $\mathfrak{g}$-valued one-form $A\in X^1$ just in 
terms of the above structures: 
 \begin{equation}
  S = \frac{1}{2} \big\langle A,dA\big\rangle +\frac{1}{3!}  \big\langle A, \big[A,A\big]\big\rangle \;.  
 \end{equation} 
Using the axioms (\ref{strictLinfty}) of a dgLa, together with cyclicity,  one quickly verifies that this theory is gauge invariant, 
with gauge transformations and  field equations, respectively, given by 
 \begin{equation}
  \delta A = d\lambda + [A,\lambda]\;, \qquad 
  F(A) := dA+\tfrac{1}{2} [A,A] = 0\;, \\
 \end{equation} 
where $\lambda\in X^0$ is the gauge parameter. Since  $F\in X^2$ defines the field equations  
we can think of $X^2$ as the `space of field equations' or, in line with the BV formalism, 
as the space of anti-fields. Furthermore, since the expression of the Bianchi or Noether identity  
$dF+[A,F]=0$ is a three-form we can view $X^{3}$ as the `space of Noether identities'. 
This algebraic interpretation extends to arbitrary gauge field theories, 
possibly with further spaces encoding gauge-for-gauge symmetries, etc., 
and  generally with a genuine $L_{\infty}$-algebra instead of a dgLa.

Returning  to the dgca that defined the dgLa as the tensor product with the color Lie algebra $\mathfrak{g}$ 
 via  (\ref{tensorbracket}) one may say that  the algebra of differential forms is  the `kinematic algebra' 
of Chern-Simons theory 
in the sense that this is what is left after `stripping off color'. 
It must be emphasized, however, that this is not the `kinematic Lie algebra' of amplitudes.
 In the remainder of this section we will  
 uncover the latter, following and generalizing   \cite{Ben-Shahar:2021zww}.

\subsection{BV algebra and kinematic Lie algebra} 

The additional structure needed to identify  the kinematic Lie algebra 
only reveals itself once we  give up the manifest topological invariance of Chern-Simons theory 
by introducing a fiducial metric $g_{\mu\nu}$, 
as indeed is necessary whenever one performs quantization and gauge fixing.  
Given such a metric, which we assume to be of Lorentzian signature $(-,+,+)$, one has the Hodge duality operation 
$\star: \Omega^{p}\rightarrow \Omega^{3-p}$, in terms of which one can define the adjoint $d^{\dagger}$
to the de Rham differential. Defining the inner product on $p$-forms  $u_1, u_2\in \Omega^p$: 
 \begin{equation}\label{innerFormproduct} 
  (u_1, u_2) := \int u_1\wedge \star u_2\,, 
 \end{equation}
one demands, for a $(p-1)$-form $u$ and a $p$-form $v$, that 
 \begin{equation}\label{daggerDef}
  (du ,v) = (u ,d^{\dagger} v)\;. 
 \end{equation}
From this definition it follows that $d^{\dagger}$ \textit{decreases} the form degree by one 
and is also nilpotent: $(d^{\dagger})^2=0$. Thus, the de Rham complex carries now a second 
`differential', whose degree is opposite to that of $d$. Using $\star^2=-1$ one finds the explicit 
expression $d^{\dagger}u= (-1)^u \star d\, \star u$ for any form $u$. 
In line with the notation of later sections we also denote  
  \begin{equation}
   b:= -d^{\dagger}\;, 
  \end{equation}
because this obeys the same relations as the `$b-$ghost' in string field theory. In particular, 
this operator of instrinsic degree $|b|=-1$ obeys $b^2=0$ 
and anticommutes with   $d$ into the d'Alembert operator:
 \begin{equation}\label{ddBox}
  \{d,b\} = - dd^{\dagger} -d^{\dagger}d = \square\;. 
 \end{equation} 

Given the second differential $b=-d^{\dagger}$ we can ask whether it acts as a derivation, i.e., whether it 
obeys a Leibniz rule with respect to $m_2$ (the wedge product). This turns out not to be the case. 
Rather, the \textit{failure} of $b$ to act as a derivation defines an interesting new structure: 
Setting 
 \begin{equation}\label{CSb2}
  b_2(u_1,u_2) := (-1)^{u_1}\big(bm_2(u_1,u_2) -m_2(bu_1, u_2) -  (-1)^{u_1}m_2(u_1, bu_2) \big) \,, 
 \end{equation}
one obtains a degree $-1$ graded antisymmetric bracket with respect to a degree shifted by one, 
 \begin{equation}\label{gradedanti}
  b_2(u_1,u_2) = -(-1)^{(u_1+1)(u_2+1)} b_2(u_2,u_1)\;, 
 \end{equation}
that furthermore obeys a 
graded Jacobi identity and Leibniz rule. More precisely, with the same degree-one shift we 
have a Leibniz rule of 
the form  
 \begin{equation}\label{BVLeibniz} 
  b(b_2(u_1,u_2)) = b_2(bu_1, u_2) +(-1)^{u_1+1} b_2(u_1, bu_2) \;, 
 \end{equation} 
which follows quickly just using the definition (\ref{CSb2}) and $b^2=0$.
Moreover, we have the graded Jacobi identity 
 \begin{equation}\label{gradedJacobi}
  b_2(b_2(u_1,u_2), u_3)+(-1)^{(u_1+1)(u_2+u_3)} b_2(b_2(u_2,u_3), u_1) 
  +(-1)^{(u_3+1)(u_1+u_2)} b_2(b_2(u_3,u_1), u_2) = 0 \,,
 \end{equation} 
and  a compatibility condition between $m_2$ and $b_2$: 
 \begin{equation}\label{BVcompatibility} 
  b_2(u_1, m_2(u_2, u_3)) = m_2(b_2(u_1,u_2), u_3) + (-1)^{(u_1+1)u_2} m_2(u_2, b_2(u_1,u_3))\;. 
 \end{equation}
These two relations are quite non-trivial and have to be verified by explicit computations 
using the wedge product $\wedge $ and $d^{\dagger}$.

The above is an example of a \textit{Batalin--Vilkovisky algebra} (or BV algebra for short):  
This is a  graded vector space with a degree-$(-1)$ 
differential $b$ obeying $b^2=0$ (a chain complex) 
equipped with a graded commutative and  associative product $m_2$, and a 
differential  graded Lie algebra structure with differential $b$ and Lie bracket $b_2$
satisfying the compatibility condition (\ref{BVcompatibility}) between 
$m_2$ and $b_2$. [Upon ignoring the differential, a BV algebra is known as 
Gerstenhaber algebra, which is a generalization of the Poisson algebra of functions 
on phase space. Here the product is just the ordinary product of functions and the Lie bracket is 
the Poisson bracket, which indeed satisfies the compatibility condition (Poisson identity).]

This definition, as well as the explicit check of the two relations (\ref{gradedJacobi}), (\ref{BVcompatibility}), 
can be simplified by noting that in a BV algebra the differential is of `second order'. 
To explain this notion for our special case note that while $d^{\dagger}$ is defined 
in terms of a first-order differential operator it does not act via the Leibniz rule on the wedge product, 
as noted above, and in this sense is of higher order.
It is actually of second order in that it 
acts like the Laplacian on a product of functions.\footnote{More precisely, $d^{\dagger}$ being of second order means 
\begin{equation}
 d^{\dagger} (uvw) = -(d^{\dagger}u)vw - (-1)^u u(d^{\dagger}v)w-(-1)^{u+v}uvd^{\dagger}w
 +d^{\dagger}(uv)w+(-1)^u u d^{\dagger}(vw)+(-1)^{(u+1)v} vd^{\dagger}(uw) \,,
\end{equation}
where we left the wedge product implicit. The second order character of $d^{\dagger}$ is clear in the equivalent space 
of polyvectors, see (\ref{BVLaplacian}) below.} 
One can then define a BV algebra 
as a graded commutative associative algebra equipped with a differential of second order. The graded Lie bracket 
is then a derived notion, defined as in (\ref{BVLeibniz}) as the failure of the differential to obey the Leibniz rule with respect 
to the graded commutative product. Both the Jacobi identity and the compatibility condition 
are consequences of the differential being second order.

After this abstract discussion let us return to the example at hand, 
which actually has the following  simple geometric interpretation. Given the metric we 
can identify  differential forms with polyvectors (completely anti-symmetric 
contravariant tensors) by raising indices.  The inner product (\ref{innerFormproduct}) on forms then 
gives rise to the natural pairing between a $p$-form and a rank-$p$ polyvector. 
This pairing does not depend on the full metric but only on the volume form, whose 
corresponding density we denote by $\rho=\sqrt{|g|}$.  With (\ref{daggerDef}) it then follows that 
$d^{\dagger}$  is transported  to the covariant divergence on polyvectors, which we denote by $\Delta$, 
and which indeed decreases the rank by one. 
On a rank-$p$ polyvector $u^{\mu_1\ldots \mu_p}$ we have 
 \begin{equation}
  (\Delta u)^{\mu_1\ldots \mu_{p-1}} : = {\rho}^{-1} \partial_{\nu}\big(\rho\, u^{\nu\mu_1\ldots \mu_{p-1}}\big)\;. 
 \end{equation}
This is a differential in that $\Delta^2=0$ but it does not act via the Leibniz rule on the wedge product 
of polyvectors. Rather, the failure defines the so-called Schouten--Nijenhuis bracket on polyvectors, 
which for vector fields reduces to the familiar Lie bracket generating  infinitesimal diffeomorphisms. 
Indeed, setting 
 \begin{equation}
  [u_1,u_2] := (-1)^{u_1}\big(\Delta(u_1\wedge u_2) - \Delta u_1 \wedge u_2 - (-1)^{u_1} u_1\wedge \Delta u_2\big) \;, 
 \end{equation}
and specializing to vector fields $u_1, u_2$ one finds 
 \begin{equation}
 \begin{split}
  [u_1,u_2]^{\mu} &= -\big(\rho^{-1}\partial_{\nu}\big(\rho\,2 u_1^{[\nu} u_2^{\mu]}\big) -\rho^{-1}\partial_{\nu}(\rho u_1^{\nu}) u_2^{\mu}
  +u_1^{\mu} \rho^{-1}\partial_{\nu} (\rho u_2^{\nu})\big) \\
  &= u_2^{\nu}\partial_{\nu} u_1^{\mu} - u_1^{\nu}\partial_{\nu} u_2^{\mu}\;, 
 \end{split} 
 \end{equation} 
which is the diffeomorphism covariant Lie bracket of vector fields (in which the volume factors 
have cancelled). This bracket, and the Schouten--Nijenhuis bracket more generally, of course satisfy 
the Jacobi identity. Moreover, the compatibility condition (\ref{BVcompatibility}) has a simple geometric interpretation: 
it means that the wedge product of polyvectors is covariant under infinitesimal diffeomorphisms. 
Thus, the polyvectors equipped with the wedge product and the second order differential $\Delta$ 
form a BV algebra.

As an aside, let us note that in this picture of polyvectors there is a particularly intuitive way to 
understand that $\Delta$ is of second order and hence defines a BV algebra. Following \cite{Witten:1990wb} 
we start by viewing the de Rham complex as functions of even coordinates $x^{\mu}$ and odd anti-commuting coordinates 
$\theta^{\mu}$ playing the role of $dx^{\mu}$. The expansion of a function $f(x,\theta)$ reads 
 \begin{equation}
  f(x,\theta) = \sum_p\tfrac{1}{p!}\, f_{\mu_1\ldots \mu_p}(x)\, \theta^{\mu_1}\cdots \,\theta^{\mu_p} \,, 
 \end{equation}
and thus this space of functions is equivalent to the de Rham complex of differential forms. Moreover, the pointwise product $f\cdot g$  of 
functions encodes the wedge product of differential forms. The de Rham differential is now realized as 
 \begin{equation}
  d = \theta^{\mu}\frac{\partial}{\partial x^{\mu}}\,, 
 \end{equation}
and thus, taking the form of a vector field, acts as a derivation on the product. 
Turning then to the chain complex of polyvector fields, these can be realized as functions of $x^{\mu}$ and new odd variables 
$\vartheta_{\mu}$, 
 \begin{equation}
  F(x,\vartheta) = \sum_p \tfrac{1}{p!} \, F^{\mu_1\ldots \mu_p}(x)  \, \vartheta_{\mu_1}\cdots \,\vartheta_{\mu_p}\,, 
 \end{equation}
for which the pointwise product yields the wedge product of polyvectors.  The differential given by 
the above divergence operator $\Delta$ is then realized, say for trivial volume measure $\rho=1$,  as 
 \begin{equation}\label{BVLaplacian} 
  \Delta = \frac{\partial^2}{\partial x^{\mu} \partial \vartheta_{\mu}} \,.
 \end{equation}
This  makes it manifest that $\Delta$ is of  second order with respect to the wedge product of polyvectors and hence, in the isomorphic space of differential forms, 
that  $d^{\dagger}$ is second order. 

After this aside,  we finally  turn to the  kinematic Lie algebra of Chern-Simons theory, 
which has recently been identified by 
Ben-Shahar and Johansson \cite{Ben-Shahar:2021zww} and turns out to be 
a small subalgebra of the above BV algebra. To see this we specialize to  the fields of Chern-Simons theory 
and impose the condition 
 \begin{equation}
  bA=0\,, 
 \end{equation}
which  means that  the divergence of the corresponding vector field  vanishes. 
This is just a standard gauge fixing condition (as one needs to impose  for any quantum computations). 
The Lie bracket is closed on divergence-free vector fields for which it is   known 
as the algebra of volume preserving diffeomorphisms, which  was identified in \cite{Ben-Shahar:2021zww} as 
the kinematic Lie algebra of Chern-Simons theory.  
The operator $b$ is perfectly suited to impose a gauge fixing condition, but we see here that 
there is a rich algebraic structure whether one imposes $bA=0$ or not.

We close this section by  pointing  out that  there is actually more structure than  a 
BV algebra, because the de Rham differential $d$ plays no role in the latter. A BV algebra equipped with 
a second differential (of opposite degree to the first) that acts as a derivation on the product is known as 
a differential graded BV algebra provided both differentials anti-commute. 
Here, however, they anticommute to the d'Alembert operator $\square$, see (\ref{ddBox}). Following 
Reiterer we will refer to such a structure as a BV$^{\square}$-algebra.\footnote{It was also noted in \cite{Borsten:2022vtg} that the ${\rm BV}^\B$ algebra of Reiterer is present in Chern-Simons theory.} 
While for the considerations in \cite{Ben-Shahar:2021zww} all this extra structure was not needed, this changes for genuine  Yang-Mills theory  in arbitrary dimensions. 
At least in its known local formulations, 
in order to double copy Yang-Mills theory the full 
BV$^{\square}$-algebra in its homotopy version, denoted BV$^{\square}_{\infty}$ in the following, is needed.

\section{Color-stripped  Yang-Mills theory and BV$_{\infty}^{\square}$} 

Pure Yang-Mills theory can be described, as any classical field theory, by an $L_\infty$ algebra \cite{Hohm:2017pnh,Zeitlin:2008cc}. Since all fields and parameters take values in the Lie algebra $\mathfrak{g}$ of the color gauge group, this $L_\infty$ algebra is given by the tensor product $\cK\otimes\mathfrak{g}$, where $\cK$ is the color-stripped space containing local fields and parameters with no color degrees of freedom. It was shown in \cite{Zeitlin:2008cc} that $\cK$ is endowed with a $C_\infty$ algebra structure. 
In this section we will show that $\cK$ carries a much larger algebraic structure, named ${\rm BV}_\infty^\B$ algebra in \cite{Reiterer:2019dys}, up to three arguments. 
This algebra is a vast generalization of the BV algebra associated to Chern-Simons theory \cite{Ben-Shahar:2021zww}, and is the backbone for constructing double field theory to quartic order via double copy.

\subsection{The $C_\infty$ algebra of Yang-Mills}

We employ a formulation of Yang-Mills with an auxiliary scalar field $\varphi$, which only enters in the free theory:
\begin{equation}\label{YMaction}
S=\int d^Dx\,\Big[\tfrac12\,A^\mu_a\B A_\mu^a-\tfrac12\,\varphi_a\varphi^a+\varphi_a\,\del^\mu A_\mu^a-f_{abc}\,\del_\mu A_\nu^a A^{\mu b}A^{\nu c}-\tfrac14\,f^e{}_{ab}f_{ecd}\,A_\mu^a A_\nu^b A^{\mu c}A^{\nu d}\Big]\;.    
\end{equation}
The cubic and quartic vertices are the standard ones, and integrating out $\varphi$ one recovers the usual Yang-Mills action. This form of the action was derived in \cite{Bonezzi:2022yuh} from a worldline theory, which shares some general features with open string field theory, and used to construct double field theory to cubic order.

The $C_\infty$ algebra of Yang-Mills is the graded vector space $\cK=\bigoplus_{i=0}^3K_i$, endowed with a nilpotent differential\footnote{In our previous paper \cite{Bonezzi:2022yuh} this was also denoted by $Q$, as it is the BRST operator of a suitable worldline theory.} $m_1$ of degree $+1$, together with bilinear and trilinear products $m_2$ of degree zero and $m_3$ of degree $-1$. 
The spaces of (color-stripped) gauge parameters $\lambda$, fields $\cA=(A_\mu, \varphi)$, field equations $\cE=(E_\mu, E)$ and Noether identities $\cN$ are organized in the chain complex $(\cK, m_{1})$ as follows:
\begin{equation}\label{K}
\begin{tikzcd}[row sep=2mm]
&K_{0}\arrow{r}{m_{1}} & K_{1}\arrow{r}{m_{1}} & K_2\arrow{r}{m_{1}} & K_3\\
\cK^{(0)}:&\lambda& A_\mu & E\\
\cK^{(1)}: & &\varphi&E_\mu &\cN\;,
\end{tikzcd}
\end{equation}
which shows that $\cK$ can also be decomposed as the direct sum of two isomorphic spaces:
\begin{equation}\label{Kdecomp}
\cK=\cK^{(0)}\oplus\cK^{(1)}\;,\quad (\lambda\,,\, A_\mu\,,\,E)\in\cK^{(0)}\;,\quad (\varphi\,,\, E_\mu\,,\,\cN)\in\cK^{(1)}\;.   
\end{equation}  
This decomposition, discussed in more detail in \cite{Bonezzi:2022yuh}, defines an inner product on $\cK$ as a degree $-3$ pairing between $\cK^{(0)}$ and $\cK^{(1)}$ given by
\begin{equation}\label{innerYM}
\left\l\cA,\cE\right\r:=\int d^Dx\,\big(A_\mu E^\mu+\varphi\,E\big)\;,\quad \left\l\lambda,\cN\right\r:=\int d^Dx\,\big(\lambda\,\cN\big)   \;. 
\end{equation}
Upon tensoring with $\mathfrak{g}$, the action \eqref{YMaction} can be written as
\begin{equation}
S=\tfrac12\,\big\l\cA^a,m_{1}(\cA_a)\big\r+\tfrac16\,f_{abc}\,\big\l\cA^a,m_2(\cA^b,\cA^c)\big\r+\tfrac14\,f^e{}_{ab}f_{ecd}\,\big\l\cA^a,m_{3h}(\cA^b,\cA^c|\cA^d)\big\r
\end{equation}
in terms of the $C_\infty$ inner product, which we will use in the next section to express the four-gluon amplitude.

The differential $m_{1}$ encodes the free dynamics in terms of linear field equations $m_{1}(\cA)=0$, linearized gauge transformations $\delta\cA=m_{1}(\lambda)$ and Noether identities $m_{1}(m_{1}(\cA))=0$, and is explicitly realized as
\begin{equation}\label{Qaction}
\begin{split}
m_{1}(\lambda)&=\bpm\del_\mu\lambda\\\square\lambda\epm\in K_{1}\;,\\ 
m_{1}\bpm A_\mu\\\varphi\epm&=\bpm\del\cdot A-\varphi\\\square A_\mu-\del_\mu\varphi\epm\in K_2\;, \\ 
m_{1}\bpm E\\ E_\mu\epm&=\square E-\del^\mu E_\mu\in K_3\;,  
\end{split} 
\end{equation}
and $m_{1}(\cN)\equiv0$ by degree.
The products $m_2$ and $m_3$, evaluated on fields, correspond to the color-stripped cubic and quartic vertices, respectively:
\begin{equation}
\begin{split}
m^\mu_2(A_1, A_2)&=\del\cdot A_1 A_2^\mu+2\,A_1\cdot\del A_2^\mu+\del^\mu A_1\cdot A_2 - (1\leftrightarrow2)\;,\\
m_3^\mu(A_1,A_2,A_3)&=A_1\cdot A_2\,A_3^\mu+A_3\cdot A_2\,A_1^\mu-2\,A_1\cdot A_3\,A_2^\mu\;,
\end{split}    
\end{equation}
where we use the shorthand notation $A_i\cdot A_j=A_i^\mu A_{j\mu}$ and $A\cdot\del=A^\mu\del_\mu$.
Notice that both belong to $K_2$, which is the space of equations of motion.
The non-vanishing products between arguments other than fields are given in appendix \ref{AppYM}. They encode, for instance, the nonlinear part of the gauge transformations and the gauge algebra. For a detailed discussion of the $C_\infty$ algebra of Yang-Mills we refer to \cite{Bonezzi:2022yuh}. In the following we will use $u_1,u_2,\ldots$ to  denote generic elements of the vector space $\cK$.

The symmetry property of $C_\infty$ products $m_n$ is determined by requiring that they vanish on shuffles. With our degree conventions this  reads
\begin{equation}
\begin{split}
m_2(u_1,u_2)-(-)^{u_1u_2}m_2(u_2,u_1)&=0\;,\\
m_3(u_1,u_2,u_3)-(-)^{u_1u_2}m_3(u_2,u_1,u_3)+(-)^{u_1(u_2+u_3)}m_3(u_2,u_3,u_1)&=0\;,    
\end{split}
\end{equation}
which for $m_2$ is the same as being graded symmetric.
The nontrivial $C_\infty$ relations amount to nilpotency of $m_{1}$, the Leibniz property of $m_{1}$ with respect to $m_2$ ($m_{1}$ is a derivation for $m_2$), and associativity of $m_2$ up to homotopy:
\begin{equation}\label{C-relations}
\begin{split}
m_{1}^2(u)&=0\;,\\ 
m_{1}(m_2(u_1,u_2))&=m_2(m_{1}(u_1),u_2)+(-)^{u_1}m_2(u_1, m_{1}(u_2))\;,\\ 
m_2\big(m_2(u_1,u_2), u_3\big)-m_2\big(u_1,m_2(u_2,u_3)\big)&=m_{1}(m_3(u_1,u_2,u_3))+m_3(m_{1}(u_1),u_2,u_3)\\
&+(-)^{u_1}m_3(u_1,m_{1}(u_2),u_3)+(-)^{u_1+u_2}m_3(u_1,u_2,m_{1}(u_3)) \;.
\end{split}    
\end{equation}
Nilpotency of the differential ensures gauge invariance of the linearized theory, while the Leibniz and associativity relations encode consistency of the color-stripped cubic and quartic interactions, respectively.
For our purposes, it will be
useful to redefine the $m_3$ product as
\begin{equation}
\begin{split}
m_{3h}(u_1,u_2|u_3)&:=\tfrac13\,\Big(m_3(u_1,u_2,u_3)+(-)^{u_1u_2}m_3(u_2,u_1,u_3)\Big)\;,\\
m_{3}(u_1,u_2,u_3)&=m_{3h}(u_1,u_2|u_3)-(-)^{u_1(u_2+u_3)}m_{3h}(u_2,u_3|u_1)\;.
\end{split}    
\end{equation}
Let us emphasize that $m_{3h}$ contains precisely the same information as $m_3$, and the latter can indeed be reconstructed from the former. 
The redefined product $m_{3h}$ is a graded hook in the labels, meaning that it is graded symmetric in the first two inputs (which we highlight by the vertical bar) and vanishes upon total graded symmetrization. 
Evaluated on Yang-Mills fields it reads
\begin{equation}
m_{3h}^\mu(\cA_1,\cA_2|\cA_3)=A_1^\mu A_2\cdot A_3-A_2^\mu A_1\cdot A_3\;.    
\end{equation}

\paragraph{The $b$ operator} 

It turns out that the $C_\infty$ algebra structure of Yang-Mills is not enough to construct a gravitational theory on the doubled space $\cK\otimes\bar\cK$. The main additional ingredient is a second nilpotent differential of degree $-1$, which we denote by $b$. In this formulation of Yang-Mills, the $b$ operator is the $b-$ghost of the underlying worldline theory, and acts as a local operator without spacetime derivatives:
\begin{equation}\label{baction}
b\bpm A_\mu\\\varphi\epm=\varphi\in K_{0}\;,\quad b\bpm E\\E_\mu\epm=\bpm E_\mu\\0\epm\in K_{1}\;,\quad b\cN=\bpm\cN\\0\epm\in K_2\;,    
\end{equation}
with $b\lambda\equiv0$ by degree counting. This can be visualized on the complex \eqref{K} as
\begin{equation}\label{bdiagram}
\begin{tikzcd}[row sep=2mm]
&K_{0}\arrow{r}{m_{1}} & K_{1}\arrow{r}{m_{1}} & K_2\arrow{r}{m_{1}} & K_3\\
\cK^{(0)}:&\lambda& A_\mu & E\\
\cK^{(1)}: & &\arrow{ul}{b}\varphi&\arrow{ul}{b}E_\mu &\arrow{ul}{b}\cN
\end{tikzcd}\;,
\end{equation}
with $b$ mapping from $\cK^{(1)}$ to ${\cK^{(0)}={\rm ker}(b)}$.  
More generally, the defining properties of $b$ are nilpotency and the commutation relation with the differential $m_{1}$:
\begin{equation}\label{hodge}
b^2=0\;,\quad m_{1}\,b+b\,m_{1}=\B\;. 
\end{equation}
The latter defines a generalized Hodge decomposition of the wave operator $\B=\del^\mu\del_\mu$, which has degree zero and commutes with both $m_{1}$ and $b$. In this respect $b$ can provide both a gauge fixing condition, as $b\cA=0$, and the propagator, as $\frac{b}{\B}$ acting on the space of equations (or sources).

We will now show that it is the interplay of the $b$ differential with the $C_\infty$ algebra to give rise to a much richer algebraic structure on $\cK$. The $b$ operator, in fact, is a generalization of a BV Laplacian.
In a standard BV algebra, the odd Laplacian is a nilpotent second order differential, and its failure to be a derivation of the product defines the BV antibracket. 
In the present case, one can similarly define a degree $-1$ bracket $b_2$ on $\cK$ by the failure of $b$ to be a derivation of the $C_\infty$ product $m_2$:
\begin{equation}\label{b2}
b_2(u_1,u_2):=bm_2(u_1,u_2)-m_2(bu_1,u_2)-(-)^{u_1}m_2(u_1,bu_2)\;.    
\end{equation}
Here we changed sign conventions from the previous section (see \eqref{CSb2}) to make $b_2$ graded symmetric: $b_2(u_1,u_2)=(-)^{u_1u_2}b_2(u_2,u_1)$, which is more conventional for graded Lie brackets and $L_\infty$ algebras.
In an ordinary BV algebra, the second order nature of the Laplacian is reflected by the compatibility of the antibracket with the product. Conversely, we can define $b$ to be second order with respect to $m_2$ \emph{if} the bracket $b_2$ is compatible with $m_2$, \emph{i.e.} if the graded Poisson identity
\begin{equation}\label{Poisson}
b_2\big(v,m_2(u_1,u_2)\big)-m_2\big(b_2(v,u_1),u_2\big)-(-)^{u_1u_2}m_2\big(b_2(v,u_2),u_1\big)=0\;,    
\end{equation}
holds.
This is the case for Chern-Simons theory, as we have reviewed in the previous section, but not for Yang-Mills, at least in any standard formulation. 
In fact, the two main departure points compared to the Chern-Simons case are:
\begin{itemize}
\item $b$ is \emph{not} second order with respect to $m_2$,
\item $m_2$ is \emph{not} associative.
\end{itemize}
Both these generalizations add layers of complexity to the resulting structure, ultimately leading to the concept of ${\rm BV}_\infty^\B$ \cite{Reiterer:2019dys}. In appendix \ref{AppBV} we describe how a BV algebra can be generalized by relaxing its axioms step by step, and ignoring the $\B$ operator, in order to convey the logic in a simpler setup.
Before proceeding further, we will introduce a more streamlined notation for dealing with relations of maps and operators in $\cK$.

\subsection{Intrinsic input-free formulation}

In order to simplify our construction, it will be useful to formulate all the algebraic relations as input-free equations between maps. We shall start by defining the commutator between an operator $\cO:\cK\rightarrow\cK$ and a bilinear map $\cM$ as the bilinear map $[\cO,\cM]$ acting as
\begin{equation}\label{OMcommutator}
[\cO,\cM](u_1, u_2):=\cO\cM (u_1,u_2)-(-)^{|\cO||\cM|}\Big[\cM(\cO u_1,u_2)+(-)^{u_1|\cO|}\cM(u_1,\cO u_2)\Big]\;.   
\end{equation}
Similarly, the commutator with a trilinear map $\cT$ is the trilinear map $[\cO,\cT]$ given by
\begin{equation}\label{OTcommutator}
\begin{split}
[\cO,\cT](u_1, u_2, u_3)&:=\cO\cT(u_1, u_2,u_3)-(-)^{|\cO||\cT|}\Big[\cT(\cO u_1,u_2,u_3)\\
&+(-)^{|\cO|u_1}\cT(u_1,\cO u_2,u_3)+(-)^{|\cO|(u_1+u_2)}\cT(u_1,u_2,\cO u_3)\Big]\;.    
\end{split}    
\end{equation}
With these definitions one can show that multiple commutators obey 
\begin{equation}
[\cO_1,[\cO_2,\cM]]-(-)^{|\cO_1||\cO_2|}[\cO_2,[\cO_1,\cM]]=[[\cO_1,\cO_2],\cM]\;,    
\end{equation}
where we used the standard graded commutator of operators
\begin{equation}
[\cO_1,\cO_2](u):=\cO_1\big(\cO_2 u\big)-(-)^{|\cO_1||\cO_2|}\cO_2\big(\cO_1 u\big)\;,    
\end{equation}
the same holding for the nested commutator $[\cO_1,[\cO_2,\cT]]$ with a trilinear map.
Next, we introduce the composition of two bilinear maps $\cM_1$ and $\cM_2$ as the trilinear map $\cM_1\cM_2$ defined by the nesting from the left:
\begin{equation}
\cM_1\cM_2(u_1, u_2, u_3):=\cM_1\big(\cM_2(u_1, u_2), u_3\big)\;.    
\end{equation} 
This is sufficient for our purposes, due to the graded symmetry of all bilinear maps involved. 
One can then prove that the commutator with an operator $\cO$ distributes according to
\begin{equation}
[\cO, \cM_1\cM_2]=[\cO, \cM_1]\cM_2+(-)^{|\cO||\cM|} \cM_1[\cO,\cM_2]\;.    
\end{equation}
It is important to notice that the left-hand side above is the commutator of $\cO$ with the trilinear map $\cM_1\cM_2$, while the right-hand side is given by the composition of the bilinear maps $\cM_{i}$ and $[\cO,\cM_{j}]$.
In order to deal with cyclic sums, we introduce a degree zero operator $\Delta$, which performs cyclic permutations of three inputs:
\begin{equation}\label{DeltaCyc}
\Delta(u_1, u_2, u_3):=(-)^{u_1(u_2+u_3)}(u_2, u_3, u_1)\;.    
\end{equation}
Its repeated action yields $\Delta^2(u_1, u_2, u_3)=(-)^{u_3(u_1+u_2)}(u_3, u_1, u_2)$ and $\Delta^3=1$. One can then define the cyclic invariant projector $\pi$ by
\begin{equation}
\pi:=\tfrac13\,\big(1+\Delta+\Delta^2\big)\;,\quad \pi^2=\pi\;,\quad \pi\Delta=\pi\;,    
\end{equation}
which decomposes the three-input space along the orthogonal subspaces generated by $\pi$ and $1-\pi$. Since every trilinear map $\cT$ to be considered in the following is graded symmetric in the first two entries,\footnote{This is the case because it can only be determined by the left nesting $\cM_1\cM_2$ of two graded symmetric bilinear maps, or by $m_{3h}$.} the projections by $\pi$ and $1-\pi$ coincide with the projections onto the totally graded symmetric and graded hook parts $\cT_s$ and $\cT_h$, respectively:
\begin{equation}\label{symhook}
\cT=\cT_s+\cT_h\;,\quad
\cT_s:=\cT\pi\;,\quad\cT_h:=\cT(1-\pi)\;.    
\end{equation}
With graded hook we refer to the representation that 
vanishes upon total graded symmetrization, so that 
$\cT_h\pi=0$. 
The projection by $\pi$ (or the action of $\Delta$)  commutes with any operator $\cO$, in the sense that
\begin{equation}
[\cO,\cT]\Delta=[\cO,\cT\Delta]\;,    
\end{equation}
as it can be checked directly from the definitions \eqref{OTcommutator} and \eqref{DeltaCyc}.

Among the operators $\cO:\cK\rightarrow\cK$, the wave operator $\B$ plays an important role, in that it is responsible for the more exotic deformations of the algebraic relations. According to the definition \eqref{OMcommutator}, its commutator with a bilinear map $\cM$ is given by
\begin{equation}
[\B,\cM](u_1,u_2)=2\,\cM(\del^\mu u_1,\del_\mu u_2)\;.
\end{equation}
In order to treat the possible actions of $\B$ on three arguments, we shall further define
\begin{equation}\label{dboxes}
\begin{split}
d_s(u_1, u_2, u_3)&:=2\,(\del^\mu u_1,\del_\mu u_2, u_3)\;,\\
d_t(u_1, u_2, u_3)&:=2\,(u_1,\del^\mu u_2, \del_\mu u_3)\;,\quad d_t=\Delta^2d_s\Delta\\
d_u(u_1, u_2, u_3)&:=2\,(\del^\mu u_1, u_2,\del_\mu u_3)\;,\quad d_u=\Delta d_s\Delta^2\;,\\
d_\B&:=d_s+d_t+d_u\;,
\end{split}    
\end{equation}
borrowing from the standard notation for Mandelstam invariants. With these definitions, one can see that commutators with $\B$ can be expressed as
\begin{equation}
\begin{split}
[\B,\cT]&=\cT d_\B\;,\quad\cM_1[\B, \cM_2]=\cM_1\cM_2 d_s\;,\quad [\B,\cM_1]\cM_2=\cM_1\cM_2(d_t+d_u)\;,    
\end{split}    
\end{equation}
for bilinear maps $\cM_1$, $\cM_2$ and trilinear map $\cT$. 
From \eqref{dboxes} it is easy to derive that $d_\B$ is cyclic invariant: $d_\B\Delta=\Delta d_\B$, and that
\begin{equation}\label{dboxrel}
d_\B\pi=\pi d_\B=3\,\pi d_s\pi\;,    
\end{equation}
which will be important in the following. Lastly, since both $m_{1}$ and $b$ commute with $\B$, one can see that they also commute with $d_s$ (and thus with $d_\B$), in the sense that
\begin{equation}\label{Qdscommutator}
[m_{1},\cT]d_s=[m_{1},\cT d_s]\;,\quad [b,\cT]d_s=[b,\cT d_s]\;.
\end{equation}

\subsection{Constructing ${\rm BV}_\infty^\B$}

We are now in the position to construct the ${\rm BV}_\infty^\B$ algebra associated to Yang-Mills theory, starting from its $C_\infty$ algebra and $b$ operator.
Using the input-free notation, we can express the $C_\infty$ relations \eqref{C-relations} as equations for maps:
\begin{equation}
\begin{split}
m_{1}^2&=0\;,\quad\hspace{15mm} |m_{1}|=+1\;,\\
[m_{1},m_2]&=0\;,\quad\hspace{13mm} |m_2|=0\;,\\
m_2m_2\,(1-\pi)&=[m_{1},m_{3h}]\;,\quad |m_{3h}|=-1\;, 
\end{split}    
\end{equation}
with the symmetry property of the 3-product stated as $m_{3h}\pi=0$.
Similarly, the $b$ operator and 2-bracket \eqref{b2} are given by
\begin{equation}
\begin{split}
b^2&=0\;,\quad [m_{1},b]=\B\;,\quad |b|=-1\;,\\
b_2&:=[b,m_2]\;,\quad \hspace{13mm}|b_2|=-1\;.
\end{split}
\end{equation}
If $b$ commuted with the differential $m_{1}$, the latter would be a derivation of the bracket $b_2$, since ${[m_{1},m_2]=0}$ would imply ${[m_{1}, b_2]=0}$. In a field theory (including Chern-Simons) however, the $\B$ generates further deformations of the algebraic structures. From the definition of $b_2$ it follows that
\begin{equation}
[m_{1},b_2]=[\B,m_2] \;,\quad {\rm with}\quad [\B,m_2](u_1, u_2)=2\,m_2(\del^\mu u_1,\del_\mu u_2) \;,  
\end{equation}
that is, $m_{1}$ is a derivation of the bracket \emph{up to} $\B$. 

Let us turn now to the Poisson compatibility between $b_2$ and $m_2$. We have already mentioned that \eqref{Poisson} does not hold in Yang-Mills or, which is the same, that $b$ is not second order.
We shall thus parametrize the failure of $b$ to be second order by a degree $-1$ map $k_3$, defined by
\begin{equation}\label{k3}
\begin{split}
k_3(u_1,u_2|u_3)&:=b_2\big(m_2(u_1,u_2),u_3\big)-(-)^{u_1(u_2+u_3)}m_2\big(b_2(u_2,u_3),u_1\big)\\
&\hspace{35mm}-(-)^{u_3(u_1+u_2)}m_2\big(b_2(u_3,u_1),u_2\big)\;,    
\end{split}
\end{equation}
where we rearranged the expression in \eqref{Poisson} to have a uniform nesting from the left. Notice that $k_3(u_1,u_2|u_3)$ is graded symmetric in the first two inputs. Its input-free definition is given by
\begin{equation}\label{b2k3}
k_3:=b_2m_2-m_2b_2(\Delta+\Delta^2)\;,\quad |k_3|=-1\;.    
\end{equation}
Using $b_2m_2+m_2b_2=[b,m_2m_2]$, one can rewrite $k_3$ in a more useful form:
\begin{equation}\label{k3useful}
k_3=[b,m_2m_2]-3\,m_2b_2\,\pi\;,   
\end{equation}
which shows that the hook projection $k_3(1-\pi)$ is $b-$exact and that $[b,k_3]$ is totally graded symmetric, which will be important in the following.

The natural relaxation of the compatibility \eqref{Poisson}
would be to hold up to homotopy. In terms of the failure $k_3$, this would amount to $k_3=[m_{1},\theta_3]$ for some $\theta_3$ of degree $-2$. Due to the $\B$ obstruction and the lack of associativity, however, one expects a deformation. To determine this, we shall compute the $m_{1}-$commutator of $k_3$. Using the form \eqref{k3useful} one obtains
\begin{equation}
\begin{split}
[m_{1},k_3]&= [m_{1},[b,m_2m_2]]-3\,[m_{1},m_2b_2]\,\pi\\
&=[\B,m_2m_2]-3\,m_2[\B,m_2]\,\pi=m_2m_2\,(d_\B-3\,d_s\pi)\\
&=m_2m_2\,(1-\pi)\,(d_\B-3\,d_s\pi)=[m_{1},m_{3h}]\,(d_\B-3\,d_s\pi)\\
&=[m_{1},m_{3h}\,(d_\B-3\,d_s\pi)]\;,
\end{split}    
\end{equation}
where we used homotopy associativity and \eqref{dboxrel}, \eqref{Qdscommutator}. This shows that $k_3-m_{3h}(d_\B-3\,d_s\pi)$, rather than $k_3$, ought to be $m_{1}-$exact. For Yang-Mills theory we computed explicitly $k_3$, using \eqref{baction} for $b$ and \eqref{m2s} for $m_2$, and proved that this is the case, namely that
\begin{equation}\label{homPoiss}
[b,m_2m_2]-3\,m_2b_2\,\pi=[m_{1}, \theta_3]+m_{3h}(d_\B-3\,d_s\pi)\;,
\end{equation}
with a local expression for the map $\theta_3$, which we give in appendix \ref{AppYM}. This is the deformed Poisson compatibility of the ${\rm BV}_\infty^\B$ algebra, which will play a central role in constructing double field theory at quartic order. The Poisson homotopy map $\theta_3$ is not completely independent. Its graded hook part $\theta_{3h}=\theta_3\,(1-\pi)$ is determined by the $C_\infty$ product $m_{3h}$:
\begin{equation}\label{thetahook}
\theta_{3h}+[b,m_{3h}]=0\;,   
\end{equation} 
which can be checked explicitly from the expressions \eqref{theta2} and \eqref{theta3} for $\theta_3$, while in general one can only prove that the left-hand side of \eqref{thetahook} is $m_{1}-$closed upon projecting \eqref{homPoiss} with $1-\pi$.
The symmetric projection of \eqref{homPoiss}, on the other hand, yields
\begin{equation}\label{k3s}
k_{3s}=[b,m_2m_2\pi]-3\,m_2b_2\pi=[m_{1},\theta_{3s}]-3\,m_{3h}d_s\pi \;,   
\end{equation}
upon using $[d_\B, \pi]=0$ and $m_{3h}\pi=0$.
One can use this to probe the Jacobi relation of the bracket $b_2$, since by taking a $b-$commutator of $k_{3s}$ one obtains
\begin{equation}
[b,k_{3s}]=-3\,[b,m_2b_2]\pi=-3\,b_2b_2\pi=-{\rm jac}_{b_2}\;,    
\end{equation}
where the jacobiator ${\rm jac}_{b_2}$ is defined by the graded cyclic sum
\begin{equation}
\begin{split}
{\rm jac}_{b_2}(u_1,u_2,u_3)&:=3\,b_2b_2\pi(u_1, u_2 , u_3)\\
&=b_2\big(b_2(u_1,u_2),u_3\big)+(-)^{u_1(u_2+u_3)}b_2\big(b_2(u_2,u_3),u_1\big)\\
&\hspace{33mm}+(-)^{u_3(u_1+u_2)}b_2\big(b_2(u_3,u_1),u_2\big)\;.
\end{split}    
\end{equation}
In a standard $L_\infty$ algebra, the 2-bracket $b_2$ obeys the Jacobi identity up to homotopy. In this case we expect a further deformation due to $\B$ obstructions. To show this, we use the Poisson relation \eqref{k3s} to rewrite $[b,k_{3s}]$ as follows: 
\begin{equation}
\begin{split}
[b,k_{3s}]&=[b,[m_{1},\theta_{3s}]]-3\,[b,m_{3h}]d_s\pi\\
&=-[m_{1},[b,\theta_{3s}]]+[\B, \theta_{3s}]-3\,[b,m_{3h}]d_s\pi\\
&=-[m_{1},[b,\theta_{3s}]]+\theta_{3s}d_\B+3\,\theta_{3h}d_s\pi=-[m_{1},[b,\theta_{3s}]]+\theta_{3}\big(\pi d_\B+3\,(1-\pi)d_s\pi\big)\\
&=-[m_{1},[b,\theta_{3s}]]+3\,\theta_{3}d_s\pi\;,
\end{split}    
\end{equation}
upon using \eqref{dboxrel} for the $\B$ deformations and \eqref{thetahook}.
We have thus shown that the bracket $b_2$ obeys the Jacobi identity up to homotopy and a $\B$ deformation, with a degree $-3$ graded symmetric 3-bracket $b_3$ which is completely determined by the symmetric part of $\theta_3$:
\begin{equation}
\begin{split}
{\rm jac}_{b_2}+[m_{1},b_3]+3\,\theta_{3}d_s\pi&=0\;,\\
b_3+[b, \theta_{3s}]&=0\;.
\end{split}    
\end{equation}
As an example, the 3-bracket $b_3$ acting on fields is a gauge parameter given by
\begin{equation}
b_3(\cA_{1},\cA_{2},\cA_{3})\stackrel{[123]}{=}-6\, A_{1}\cdot\del A_{2}\cdot A_{3}\, \in K_{0}\;.    
\end{equation}
It is noteworthy that this coincides with the map denoted  $h_{123}$ by Mafra and Schlotterer \cite{Mafra:2015vca}, where it plays the role of a composite gauge parameter.

This concludes the hierarchy of algebraic relations of the ${\rm BV}_\infty^\B$ algebra up to three arguments, which we summarize here for convenience:
\begin{equation}\label{BVcrazy}
\begin{array}{ll}
m_{1}^2=0\;,\quad b^2=0\;,\quad [m_{1},b]=\B\;,&\text{differentials and central obstruction,}\\[3mm]
[m_{1},m_2]=0\;,\quad m_2m_2(1-\pi)=[m_{1},m_{3h}]\;,& C_\infty\;{\rm structure},\\[3mm]
b_2=[b,m_2]\;,\quad[m_{1},b_2]=[\B,m_2]\;,& \text{two-bracket and deformed Leibniz},\\[3mm]
b_2m_2-m_2b_2(\Delta+\Delta^2)=[m_{1},\theta_3]+m_{3h}(d_\B-3\,d_s\,\pi)\;,&\text{deformed homotopy Poisson},\\[3mm]
3\,b_2b_2\pi+[m_{1},b_3]+3\,\theta_{3}d_s\pi=0\;,&\text{deformed homotopy Jacobi},\\[3mm]
\theta_{3h}+[b,m_{3h}]=0\;,\quad b_3+[b,\theta_{3s}]=0\;, & \text{compatibility of homotopies.}
\end{array}  
\end{equation}
The $C_\infty$ products $m_n$ have standard degrees $|m_n|=2-n$, while the brackets $b_n$ of the deformed $L_\infty$ structure have unconventional degrees $|b_n|=3-2n$. On the degree-shifted space $\cK[2]$ one would have $|b_n|=+1$, but we will keep $\cK$ as in \eqref{K} instead.
In the next section we will show that the ${\rm BV}_\infty^\B$ structure of Yang-Mills theory allows to construct the $L_\infty$ algebra of $N=0$ supergravity up to its three brackets, which encode the whole quartic theory.

\section{Double Copy}

In this section we will briefly review how the $L_\infty$ algebra $\cV$ of double field theory, and hence of $N=0$ supergravity, is encoded in the tensor product space $\cK\otimes\bar\cK$ \cite{Bonezzi:2022yuh}. We will revisit the differential and 2-bracket on $\cV$ in terms of Yang-Mills building blocks, before turning to the main result of the paper, which is the explicit construction of the 3-bracket of double field theory  from Yang-Mills.

\subsection{$B_2$ and $B_3$} 

In \cite{Bonezzi:2022yuh} it was shown that the $L_\infty$ algebra $\cV$ of double field theory is a subspace of $\cK\otimes\bar\cK$, where $\cK$ and $\bar\cK$ are two copies of the Yang-Mills kinematic spaces, endowed with their respective $C_\infty$ and ${\rm BV}_\infty^\B$ structures. In particular, elements of $\cK\otimes\bar\cK$ are local fields on a doubled spacetime with coordinates $(x^\mu,\bar x^{\bar\mu})$, which is a defining feature of DFT. The graded vector space $\cV$ is defined by
\begin{equation}\label{Vconstraint}
\cV=\big\{\Psi\in\cK\otimes\bar\cK\;|\;b^-\Psi=0\;,\;(\B-\bar\B)\Psi=0\big\}\;,    
\end{equation}
where $b^\pm$ are linear combinations of the $b$ operators of the two copies:
\begin{equation}
b^\pm:=\tfrac12\,(b\pm\bar b)\;,\quad \big(b^\pm\big)^2=0\;,\quad b^+b^-+b^-b^+=0\;,    
\end{equation}
and $\B=\del^\mu\del_\mu$, $\bar\B=\bar\del^{\bar\mu}\bar\del_{\bar\mu}$ are the d'Alembertians constructed with two copies of the Minkowski metric $\eta_{\mu\nu}$ and $\eta_{\bar\mu\bar\nu}$.
The constraints \eqref{Vconstraint} originate from level matching in closed string theory and, in the form \eqref{Vconstraint}, define the so-called weakly constrained DFT. In the following we will rather consider a stronger constraint, namely $\B\equiv\bar\B$ as operators. Acting on products of fields, this implies that
\begin{equation}\label{strong}
\del^\mu f\,\del_\mu g=\bar\del^{\bar\mu} f\,\bar\del_{\bar\mu} g  \;,  
\end{equation}
for any local functions $f(x,\bar x)$ and $g(x,\bar x)$. Double field theory subject to \eqref{strong} is known as strongly constrained DFT, which is essentially equivalent to $N=0$ supergravity. The standard supergravity solution of \eqref{strong} is to set $\del_\mu=\bar\del_{\bar\mu}$, which amounts to identifying the coordinates $x^\mu$ and $\bar x^{\bar\mu}$. From now on we will thus work with the smaller subspace
\begin{equation}
\cV_{\rm strong}=\big\{\Psi\in\cK\otimes\bar\cK\;|\;b^-\Psi=0\;,\;\B\equiv\bar\B\big\}\;.    
\end{equation}
It should be emphasized that the constraint $b^-\Psi=0$ removes half of the states from $\cK\otimes\bar\cK$, but leaves otherwise unconstrained fields on $\cV$ or $\cV_{\rm strong}$, since the $b-$operator \eqref{baction} does not contain spacetime derivatives. Thanks to the decomposition \eqref{Kdecomp} of $\cK$ and $\bar\cK$, one can explicitly characterize the space ${\rm ker}(b^-)$ as
\begin{equation}
{\rm ker}(b^-)=\big(\cK^{(0)}\otimes\bar\cK^{(0)}\big)\oplus b^-\big(\cK^{(1)}\otimes\bar\cK^{(1)}\big)\;, \end{equation}
which allows to construct the components of $\cV$, as was discussed in detail in \cite{Bonezzi:2022yuh}.
As an example, the gauge parameter of DFT consists of the multiplet $\Lambda=(\lambda_\mu, \bar\lambda_{\bar\mu}, \eta)$, where\footnote{The $\lambda_\mu$ gauge parameter has opposite sign compared to the conventions of \cite{Hull:2009mi}.}
\begin{equation}
\lambda_\mu=A_\mu\otimes\bar\lambda\;,\quad\bar\lambda_{\bar\mu}=\lambda\otimes\bar A_{\bar\mu}\;,\quad \eta=b^-(\varphi\otimes\bar\varphi)\;.    
\end{equation}
The emerging field content coincides with the original one introduced by Hull and Zwiebach in \cite{Hull:2009mi}: the tensor fluctuation $e_{\mu\bar\nu}=A_\mu\otimes\bar A_{\bar\nu}$ contains the graviton and the $B-$field. It is accompanied by two scalars $e$ and $\bar e$, encoding the dilaton and a pure gauge degree of freedom, and two vector auxiliaries $f_\mu$ and $\bar f_{\bar\mu}$.

In the following we will assume that 
the tensor product of the 
space of functions of coordinates $x$ and the space of 
functions of coordinates $\bar{x}$ can be identified 
with the space of functions of $(x,\bar{x})$. 
Strictly speaking this is only true for suitable function spaces, but we will see that 
the resulting local expressions are valid in general. 
We denote arbitrary elements of $\cV_{\rm strong}$ as
\begin{equation}
\Psi(x,\bar x)=u(x)\otimes\bar u(\bar x)\;,\quad u\in\cK\;,\quad \bar u\in\bar\cK\;,    
\end{equation}
leaving linear combinations implicit. 
The action of operators $\cO:\cK\rightarrow\cK$ and $\bar\cO:\bar\cK\rightarrow\bar\cK$ is extended to $\cK\otimes\bar\cK$ by
\begin{equation}
\cO\big(u\otimes\bar u\big):=(\cO u)\otimes\bar u\;,\quad \bar\cO\big(u\otimes\bar u\big):=(-)^{|\cO|u}u\otimes(\bar\cO\bar u)\;.   
\end{equation}
This gives the proper definition of $b^\pm=\frac12\,(b\pm\bar b)$ and applies to spacetime derivatives as well, \emph{e.g.} $\bar\del_{\bar\mu}\lambda_\nu=A_\nu\otimes\bar\del_{\bar\mu}\bar\lambda$. Similarly, products of bilinear and trilinear maps of $\cK$ and $\bar\cK$ act as
\begin{equation}
\begin{split}
\big(\cM\otimes\bar\cM\big)(\Psi_1,\Psi_2)&=\big(\cM\otimes\bar\cM\big)(u_1\otimes\bar u_1,u_2\otimes \bar u_2)\\
&=(-)^{u_2\bar u_1+|\bar\cM|(u_1+u_2)}\cM(u_1,u_2)\otimes\bar\cM(\bar u_1,\bar u_2)\;,\\
\big(\cT\otimes\bar\cT\big)(\Psi_1,\Psi_2,\Psi_3)&=\big(\cT\otimes\bar\cT\big)(u_1\otimes\bar u_1,u_2\otimes \bar u_2,u_3\otimes \bar u_3)\\
&=(-)^{u_2\bar u_1+u_3(\bar u_1+\bar u_2)+|\bar\cT|(u_1+u_2+u_3)}\cT(u_1,u_2,u_3)\otimes\bar\cT(\bar u_1,\bar u_2,\bar u_3)\;.
\end{split}
\end{equation}
Lastly, operators of $\cK$ commute with operators and maps of $\bar\cK$ and vice versa: 
\begin{equation}
\begin{split}
[\cO, \cM\otimes\bar\cM](\Psi_1,\Psi_2)&=\big([\cO, \cM]\otimes\bar\cM\big)(\Psi_1,\Psi_2)\;,\\
[\bar\cO, \cM\otimes\bar\cM](\Psi_1,\Psi_2)&=(-)^{|\bar\cO||\cM|}\big(\cM\otimes[\bar\cO,\bar\cM]\big)(\Psi_1,\Psi_2)\;,\\
[\cO_1,\bar\cO_2]&=0\;,
\end{split}    
\end{equation}
with analogous relations for the tensor product of trilinear maps. Notice that this is consistent with the identification $\B=\bar\B$, in that
\begin{equation}
\begin{split}
[\B, \cM\otimes\bar\cM](\Psi_1,\Psi_2)&=\big([\B, \cM]\otimes\bar\cM\big)(\Psi_1,\Psi_2)=2\, \big(\cM\otimes\bar\cM\big)(\del^\mu\Psi_1,\del_\mu\Psi_2)\\
&=2\,\big(\cM\otimes\bar\cM\big)(\bar\del^{\bar\mu}\Psi_1,\bar\del_{\bar\mu}\Psi_2)=\big( \cM\otimes[\bar\B,\bar\cM]\big)(\Psi_1,\Psi_2)\\
&=[\bar\B, \cM\otimes\bar\cM](\Psi_1,\Psi_2)\;.
\end{split}    
\end{equation}
This allows us to extend the input-free notation of the previous section to the tensor product $\cK\otimes\bar\cK$. We will now show how the ${\rm BV}_\infty^\B$ structures of $\cK$ and $\bar\cK$ induce an $L_\infty$ structure on $\cV_{\rm strong}$ up to its 3-brackets.

\paragraph{Differential and two-bracket}

We denote the $L_\infty$ brackets on $\cV_{\rm strong}$ by $B_n$. Given the differentials $m_{1}$ and $\bar m_{1}$ of the two copies of Yang-Mills, the DFT differential is the sum
\begin{equation}
B_1=m_{1}+\bar{m}_{1}\;,\quad B_1^2=0\;.    
\end{equation}
The single-copy commutators \eqref{hodge} $[m_{1},b]=\B$ and $[\bar m_{1},\bar b]=\bar\B$ imply that $B_1$ commutes with $b^-$ on $\cV_{\rm strong}$, while $b^+$ provides the Hodge decomposition:
\begin{equation}
[B_1, b^-]=0\;,\quad[B_1, b^+]=\B\;,    
\end{equation}
thanks to the identification $\B\equiv\bar\B$. This also proves that $B_1$ is well-defined as an operator $B_1:\cV_{\rm strong}\rightarrow\cV_{\rm strong}$, since it preserves ${\rm ker}(b^-)$. 

The 2-bracket of DFT was constructed in \cite{Bonezzi:2022yuh} as
\begin{equation}
B_2=-\tfrac12\,b^-\,m_2\otimes\bar m_2\;.    
\end{equation}
This form of $B_2$ makes manifest that its image is in the kernel of $b^-$, since $(b^-)^2=0$, but somewhat obscures its algebraic nature. Since $B_2$ acts on two elements of ${\rm ker}(b^-)$, the $b^-$ operator in front is the same as a $b^-$ commutator, which allows one to express $B_2$ in an equivalent form:
\begin{equation}\label{B222}
\begin{split}
B_2&=-\tfrac12\,[b^-, m_2\otimes\bar m_2]=-\tfrac14\,[b-\bar b, m_2\otimes\bar m_2]\\
&=-\tfrac14\,\Big([b, m_2]\otimes\bar m_2-m_2\otimes[\bar b,\bar m_2]\Big)\\
&=-\tfrac14\,\Big(b_2\otimes\bar m_2-m_2\otimes\bar b_2\Big)\;.
\end{split}    
\end{equation}
This representation of $B_2$ makes it more transparent that the double copy procedure substitutes the color Lie algebra $\mathfrak{g}$ of Yang-Mills with another algebraic structure of Lie type. Consistency of the cubic theory requires that $B_1$ acts as a derivation of $B_2$, which in this language is expressed as $[B_1,B_2]=0$. This is straightforward to prove by using the deformed Leibniz rule $[m_{1},b_2]=[\B,m_2]$, the $C_\infty$ relation $[m_{1},m_2]=0$ and the strong constraint:
\begin{equation}
\begin{split}
[B_1,B_2]&=-\tfrac14\,[m_{1}+\bar{m}_{1},b_2\otimes\bar m_2-m_2\otimes\bar b_2 ]\\
&=-\tfrac14\,\Big([m_{1},b_2]\otimes\bar m_2-m_2\otimes[\bar{m}_{1},\bar b_2 ]\Big)\\
&=-\tfrac14\,[\B-\bar\B,m_2\otimes\bar m_2]=0\;.
\end{split}    
\end{equation}

\paragraph{Construction of the three-bracket}

Let us now turn to the 3-bracket of DFT. 
Given the $L_\infty$ structure to cubic order, which amounts to $B_1$ and $B_2$ obeying $B_1^2=0$ and $[B_1,B_2]=0$, the next quadratic relation is the homotopy Jacobi identity obeyed by $B_2$, which defines the 3-bracket:
\begin{equation}\label{JacB2comp}
\begin{split}
    B_{2}(B_{2}(\Psi_{1},\Psi_{2}),\Psi_{3})+(-1)^{\Psi_{1}(\Psi_{2}+\Psi_{3})}B_{2}(B_{2}(\Psi_{2},\Psi_{3}),\Psi_{1})&+(-1)^{\Psi_{3}(\Psi_{1}+\Psi_{2})}B_{2}(B_{2}(\Psi_{3},\Psi_{1}),\Psi_{2})\\
    &+[B_{1},B_{3}](\Psi_{1},\Psi_{2},\Psi_{3})=0\;.
\end{split}
\end{equation}
The strategy to construct $B_3$ is to compute the jacobiator of $B_2$ in terms of the single copy maps $m_2$ and $b_2$. The ${\rm BV}_\infty^\B$ relations \eqref{BVcrazy} of the two copies will then allow us to prove that ${\rm Jac}_{B_2}$ is a $(m_{1}+\bar{m}_{1})-$commutator, thus identifying $B_3$.

In order to proceed, let us rewrite the homotopy Jacobi relation \eqref{JacB2comp} using the input free notation introduced in the previous section:
\begin{equation}\label{inputfreehomJacB2}
    \text{Jac}_{B_{2}}+[B_{1},B_{3}]=0\;,
\end{equation}
where we defined the jacobiator $\text{Jac}_{B_{2}}$ as
\begin{equation}\label{inputfreeJAC}
   \text{Jac}_{B_{2}}:=B_{2}B_{2}\, \mathcal{C}\;,
\end{equation}
in terms of the cyclic operator on $\cK\otimes\bar \cK$, defined as 
\begin{equation}
    \mathcal{C}=\big(1\otimes 1+\Delta\otimes \bar \Delta+\Delta^2\otimes \bar \Delta^{2}\big)\;.
\end{equation}
The cyclic operator $\mathcal{C}$ obeys the following relations with the single copy projectors $\pi$ and $\bar \pi$:
\begin{equation}\label{pirels}
\begin{split}
\pi\otimes 1\, \mathcal{C}&=\pi\otimes \bar\pi\, \mathcal{C}=3\, \pi\otimes \bar \pi\;,\\
1\otimes \bar \pi\, \mathcal{C}&=\pi\otimes \bar\pi\, \mathcal{C}=3\, \pi\otimes \bar \pi\;,\\
(1\otimes 1-\pi\otimes 1)\mathcal{C}&=(1\otimes 1-1\otimes \bar\pi)\mathcal{C}=(1\otimes 1-\pi\otimes 1)(1\otimes 1-1\otimes \bar\pi)\mathcal{C}\;.\\
\end{split}
\end{equation}
These relations can be proven straightforwardly by using the fact that $\pi\Delta=\pi$.

The Jacobiator can be written in terms of the single copy brackets as 
\begin{equation}
\begin{split}
    \text{Jac}_{B_{2}}&=\frac{1}{8}\, b^{-}\, m_{2}\otimes \bar m_{2}\big\{b_{2}\otimes \bar m_{2}-m_{2}\otimes \bar b_{2}\big\}\, \mathcal{C}\\
    &=\frac{1}{8}\, b^{-}\big\{ m_{2}b_{2}\otimes\bar m_{2}\bar m_{2}-m_{2}m_{2}\otimes \bar m_{2}\bar b_{2} \big\}\, \mathcal{C}\;.
\end{split}
\end{equation}
Notice that we used two equivalent versions of the 2-bracket, namely, we wrote the outermost 2-bracket in a different but equivalent way to the innermost one. This makes the computation more economical. The next step is to use the resolution of the identity to split the maps into symmetric and hook components. The decomposition leads to
\begin{equation}
    \begin{split}
    \text{Jac}_{B_{2}}&=\frac{1}{8}\, b^{-}\, \big\{3\, m_{2}b_{2}\pi\otimes \bar m_{2}\bar m_{2}\bar \pi-3\, m_{2}m_{2}\pi\otimes \bar m_{2}\bar b_{2}\bar\pi  \big\}\\
    &+\frac{1}{8}\, b^{-}\big\{ m_{2}b_{2}(1-\pi)\otimes \bar m_{2}\bar m_{2}(1-\bar \pi)-m_{2}m_{2}(1-\pi)\otimes \bar m_{2}\bar b_{2}(1-\bar\pi)\big\}\, \mathcal{C}\;,
    \end{split}
\end{equation}
where we used the properties shown in \eqref{pirels}. One can use the symmetric projection \eqref{k3s} of the homotopy Poisson relation in the first line and the homotopy associativity relation in the second to obtain
\begin{equation}
    \begin{split}
    \text{Jac}_{B_{2}}&=\frac{1}{8}\, b^{-}\, \big\{-[m_{1},\theta_{3s}]\otimes \bar m_{2}\bar m_{2}\bar \pi+m_{2}m_{2}\pi\otimes \bar [\bar{m}_{1},\bar \theta_{3s}] \big\}\\
    &+\frac{1}{8}\, b^{-}\big\{ m_{2}b_{2}(1-\pi)\otimes [\bar{m}_{1},\bar m_{3h}]-[m_{1},m_{3h}]\otimes \bar m_{2}\bar b_{2}(1-\bar\pi)\big\}\, \mathcal{C}\\
    &+\frac{1}{8}\, b^{-}\big\{ 3\, m_{3h}d_{s}\pi\otimes \bar m_{2}\bar m_{2}\pi-3\, m_{2}m_{2}\pi\otimes \bar m_{3h}\bar d_{\bar s}\bar \pi \big\}\;,
    \end{split}
\end{equation}
where the terms involving $[b,m_{2}m_{2}]$ in the homotopy Poisson relation \eqref{homPoiss} vanish due to the $b^{-}$ constraint. 

From the above Jacobiator it is not immediate how to extract a DFT differential $B_{1}=m_{1}+\bar{m}_{1}$. This makes it difficult to read-off the explicit form of $B_{3}$ in terms of the single-copy maps. However, in order to extract a differential, we can insert zeroes in the guise of Leibniz relations as
\begin{equation}
\begin{split}
    0&=\frac{1}{8}\, b^{-}\big\{ [m_{1},m_{2}m_{2}\pi]\otimes \bar\theta_{3s}-\theta_{3s}\otimes [\bar{m}_{1},\bar m_{2}\bar m_{2}] \big\}\\
    &+\frac{1}{8}\, b^{-}\big\{ m_{3h}\otimes [\bar{m}_{1},\bar m_{2}\bar b_{2}]-[m_{1},m_{2}b_{2}]\otimes \bar m_{3h} \big\}\, \mathcal{C}\\
    &+\frac{1}{8}\, b^{-}\big\{ m_{2}m_{2}d_{s}\otimes \bar m_{3h}-m_{3h}\otimes \bar m_{2}\bar m_{2}\bar d_{\bar s} \big\}\, \mathcal{C}\;.
\end{split}
\end{equation}
The first line is zero due to the Leibniz relation of the homotopy associative algebra, whereas the second and third lines are zero by virtue of the Leibniz rule modulo box of the bracket $b_{2}$. Notice, very importantly, that the terms added as the Leibniz rule modulo box of $b_{2}$ are not projected, so they contain both their symmetric and hook components. This is particularly relevant for the terms in the last line. Adding the above zero to the Jacobiator leads to
\begin{equation}
    \begin{split}
        \text{Jac}_{B_{2}}=\frac{1}{8}\, b^{-}&\Big\{ \big[m_{1}+\bar{m}_{1},m_{2}m_{2}\pi\otimes \bar \theta_{3s}\big]-\big[m_{1}+\bar{m}_{1},\theta_{3s}\otimes \bar m_{2}\bar m_{2}\bar \pi\big] \Big\}\\
        -\frac{1}{8}\, b^{-}&\Big\{ \big[ m_{1}+\bar{m}_{1},m_{2}b_{2}(1-\pi)\otimes \bar m_{3h}  \big]+\big[m_{1}+\bar{m}_{1}, m_{3h} \otimes \bar m_{2}\bar b_{2}(1-\bar\pi)\big] \Big\}\, \mathcal{C}\\
        +\frac{1}{8}\, b^{-}&\Big\{ 3\, m_{3h}d_{s}\pi\otimes \bar m_{2}\bar m_{2}\bar \pi-m_{3h}\otimes\bar m_{2}\bar m_{2}\bar d_{\bar s}\, \mathcal{C}\\
        & +m_{2}m_{2}d_{s}\otimes \bar m_{3h}\, \mathcal{C}-3\, m_{2}m_{2}\pi\otimes \bar m_{3h}\bar d_{\bar s}\bar\pi \Big\}\;,
    \end{split}
\end{equation}
where to arrive at the terms in the second line we used the last relation in equation \eqref{pirels} in order to project some of the terms involving the $m_{2}b_{2}$ structure using the fact that $m_{3h}$ contains a $(1-\pi)$ projector. Even though it is straightforward to extract the DFT differential from the first two lines, it is not obvious how to do so in the last two lines of the Jacobiator. To this end, one has to use the strong constraint, in the form $d_s=\bar d_{\bar s}$. Let us deal explicitly with the last line:
\begin{equation}
\begin{split}
    m_{2}m_{2}d_{s}\otimes \bar m_{3h}\, \mathcal{C}-3\, m_{2}m_{2}\pi\otimes \bar m_{3h}\bar d_{\bar s}\bar\pi &=\big\{m_{2}m_{2}\otimes \bar m_{3h}\bar d_{\bar s}\, -m_{2}m_{2}\pi\otimes \bar m_{3h}\bar d_{\bar s}\big\}\, \mathcal{C}\\
    &=m_{2}m_{2}(1-\pi)\otimes \bar m_{3h}\bar d_{\bar s}\, \mathcal{C}\\
    &=[m_{1},m_{3h}]\otimes \bar m_{3h}\bar d_{\bar s}\, \mathcal{C}\;.
\end{split}
\end{equation}
In the first equality we used the first relation in equation \eqref{pirels}, and to obtain the final line we used the homotopy associativity relation. Repeating the same procedure for the other term in the Jacobiator we obtain
\begin{equation}
    \begin{split}
        \text{Jac}_{B_{2}}=\frac{1}{8}\, b^{-}&\Big\{ \big[m_{1}+\bar{m}_{1},m_{2}m_{2}\pi\otimes \bar \theta_{3s}\big]-\big[m_{1}+\bar{m}_{1},\theta_{3s}\otimes \bar m_{2}\bar m_{2}\bar \pi\big] \Big\}\\
        -\frac{1}{8}\, b^{-}&\Big\{ \big[ m_{1}+\bar{m}_{1},m_{2}b_{2}(1-\pi)\otimes \bar m_{3h}  \big]+\big[m_{1}+\bar{m}_{1}, m_{3h} \otimes \bar m_{2}\bar b_{2}(1-\bar\pi)\big] \Big\}\, \mathcal{C}\\
        +\frac{1}{8}\, b^{-}&\Big\{\big[m_{1}+\bar{m}_{1},m_{3h}\otimes \bar m_{3h}\bar d_{\bar s} \big] \Big\}\, \mathcal{C}\;.
    \end{split}
\end{equation}
From this form of the Jacobiator it is possible to read-off the $B_{3}$, which is given by
\begin{equation}\label{B3final}
\begin{split}
    B_{3}=-\frac{1}{8}\, b^{-}&\Big\{\theta_{3s}\otimes \bar m_{2}\bar m_{2}\bar\pi- m_{2}m_{2}\pi\otimes \bar \theta_{3s}\\
    &+\big[m_{2}b_{2}(1-\pi)\otimes \bar m_{3h}+m_{3h}\otimes \bar m_{2}\bar b_{2}(1-\bar\pi)\big]\, \mathcal{C}\\
    &-m_{3h}\otimes \bar m_{3h}\bar d_{\bar s}\, \mathcal{C}\Big\}\;.
\end{split}
\end{equation}
The above expression can be made simpler by noticing that given that $\theta_{3s}$ and $\bar\theta_{3s}$, and $m_{3h}$ and $\bar m_{3h}$, are projected onto their symmetric and hook parts respectively, one can drop the explicit projectors of the maps that multiply them in the tensor product. This yields 
\begin{equation}\label{niceB3}
\begin{split}
    B_{3}=-\frac{1}{8}\, b^{-}\Big\{\frac{1}{3}\, &\theta_{3s}\otimes \bar m_{2}\bar m_{2}- \frac{1}{3}\, m_{2}m_{2}\otimes \bar \theta_{3s}+m_{2}b_{2}\otimes \bar m_{3h}+m_{3h}\otimes \bar m_{2}\bar b_{2}\\
    &-m_{3h}\otimes \bar m_{3h}\bar d_{\bar s}\Big\}\, \mathcal{C}\;,
\end{split}
\end{equation}
where we factored out the cyclic operator $\mathcal{C}$ at the cost of a pre-factor of $\frac{1}{3}$ in the first two terms. Similarly to the 2-bracket $B_{2}$, there is an alternative but equivalent formulation of $B_{3}$, which reads
\begin{equation}
\begin{split}
    B_{3}=\frac{1}{16}\Big\{ &\frac{1}{3}\, b_{3}\otimes \bar m_{2}\bar m_{2}+\frac{1}{3}\, \theta_{3s}\otimes \bar b_{2}\bar m_{2}+\frac{1}{3}\, \theta_{3s}\otimes \bar m_{2}\bar b_{2}\\
    &+\frac{1}{3}\, m_{2}m_{2}\otimes\bar b_{3}+\frac{1}{3}\, b_{2}m_{2}\otimes \bar \theta_{3s}+\frac{1}{3}\, m_{2}b_{2}\otimes \bar\theta_{3s}\\
    &+m_{2}b_{2}\otimes \bar \theta_{3h}-b_{2}b_{2}\otimes \bar m_{3h}+\theta_{3h}\otimes \bar m_{2}\bar b_{2}-m_{3h}\otimes \bar b_{2}\bar b_{2}\\
    &-\theta_{3h}\otimes \bar m_{3h}\bar d_{\bar s}-m_{3h}\otimes \bar \theta_{3h}\bar d_{\bar s}\Big\}\, \mathcal{C}\;.
\end{split}
\end{equation}
This map determines gauge invariant gravity in the form of double field theory up to quartic order in the action. 

\subsection{4-graviton amplitude}

Scattering amplitudes can be formulated in the language of homotopy algebras. In this subsection we review how to express the 4-point tree-level Yang-Mills scattering amplitude using algebraic building blocks. Subsequently, as a consistency check and explicit example, we compare the 4-point graviton amplitude written in terms of the DFT brackets with the 4-point amplitude obtained by means of the BCJ double copy. 

Before looking at the amplitudes it will be convenient to discuss on-shell and gauge fixing conditions from an algebraic perspective. These are implemented by imposing that the fields obey
\begin{equation}
    m_{1}(\mathcal{A})=0\;, \quad b\mathcal{A}=0\;.
\end{equation}
These equations express that $\cA$ is on-shell and subject to the gauge condition $b\cA=\varphi=0$, which is equivalent to   
\begin{equation}
   \B A_{\mu}=0, \quad \del \cdot A=0\;.
\end{equation}
In scattering amplitudes we consider all the external fields to obey the above conditions. Thus, the gauge field can be expressed as a free wave solution
\begin{equation}\label{planes}
    A_{\mu}(x)=\epsilon_\mu(x)\,\otimes t\;,\quad\epsilon_\mu(x)=\epsilon_{\mu}(p)\, e^{ip\cdot x}\;,
\end{equation}
where $t$ is an element of the color Lie algebra, $\epsilon_{\mu}(p)$ is the polarization vector and $\epsilon_\mu(x)$ is the color-stripped gluon field. When computing amplitudes we assign  to each external particle a label $i$, a Lie algebra element $t_{i}$, and a polarization vector $\epsilon_{i\, \mu}(p_{i})$ that only depends on the momentum of said particle. The gauge fixing and on-shell conditions imply, in momentum space, that the polarization vectors and particle momenta are subject to 
\begin{equation}\label{condi}
    p_{i}\cdot \epsilon_{i}=0\;,\quad p^{2}_{i}=0\;.
\end{equation}

Let us now turn to the 4-point tree-level Yang-Mills amplitude. In terms of the algebraic maps the amplitude can be written as 
\begin{equation}\label{algebraicamp}
    \mathcal{A}_{\text{Tree}}^{(4)}=-g^{2}_{\text{YM}}\Big\langle \epsilon_{4}, \left\{m_{2}(h\, m_{2}(\epsilon_{1},\epsilon_{2}),\epsilon_{3})-m_{3h}(\epsilon_{1},\epsilon_{2}|\epsilon_{3})\right\} \Big\rangle_{\text{YM}}\, \Tr\big(t_{4}[[t_{1},t_{2}],t_{3}]\big)+\text{cyclic},
\end{equation}
where we have reinstated the Yang-Mills coupling constant, the bracket $[\cdot,\cdot]$ is the Lie bracket of the color Lie algebra, the inner product $\langle \cdot ,\cdot \rangle_{\text{YM}}$ is the inner product defined in equation \eqref{innerYM}, and we take the cyclic sum of the labels $(123)$, while keeping the label $4$ fixed. The map $m_{2}$ is the kinematic part of the cubic vertex of Yang-Mills, whereas $m_{3h}$ is the kinematic part of the quartic vertex, and we emphasize that they take the plane waves $\epsilon_i^\mu(x)$ in \eqref{planes} as inputs. The propagator $h$ is given by $h=-\frac{b}{s_{ij}}$, where $s_{ij}$ are the kinematic invariants\footnote{In our conventions, for massless particles we use: $s_{12}=s=2\, p_{1}\cdot p_{2}$, $s_{23}=t=2\, p_{2}\cdot p_{3}$, $s_{13}=u=2\, p_{1}\cdot p_{3}$.} defined as $s_{ij}=(p_{i}+p_{j})^{2}$, and the particles $i$ and $j$ are the inputs of the 2-bracket on which $h$ acts. It should be noted that since we are working in momentum space, the inner product generates momentum-conserving delta functions, i.e
\begin{equation}
    \big\langle \e_{i},J_{j} \big\rangle_{\text{YM}}=\delta^{(D)}(p_{i}+p_{j})\, \e_{i\, \mu}(p_{i})\, J^{\mu}_{j}(p_{j})\;,
\end{equation}
where $J^\mu$ denotes a current, built from external particles data, belonging to the space $K_2$ of field equations.
In the following we will not write the delta functions explicitly.

In the double copy literature Yang-Mills scattering amplitudes are usually expressed in terms of so-called kinematic numerators $n_{s{ij}}$, which depend on polarization vectors and momenta, and color factors, which are color-traces of generators of the gauge group. Explicitly, the 4-gluon amplitude can be written as
\begin{equation}\label{forbidden}
    \mathcal{A}_{\text{Tree}}^{(4)}=g^{2}_{\text{YM}}\Big\{\frac{n_{s}\, c_{s}}{s}+\frac{n_{t}\, c_{t}}{t}+\frac{n_{u}\, c_{u}}{u}\Big\}\;.
\end{equation}
Comparing equations \eqref{forbidden} and \eqref{algebraicamp}, it is possible to read-off the algebraic form of the kinematic numerators and the color factors
\begin{equation}\label{kinnum}
\begin{split}
    n_{s}&:=\big\langle \e_{4}, \mathfrak{n}_{s}\big\rangle_{\text{YM}}\;,\\
    c_{s}&:=\Tr\big(t_{4}\, [[t_{1},t_{2}],t_{3}]\big)\;,
\end{split}
\end{equation}
where we have defined the current
\begin{equation}
    \mathfrak{n}^{\mu}_{s}:= m^{\mu}_{2}(b_{2}(\epsilon_{1},\epsilon_{2}),\epsilon_{3})+s\, m_{3h}^{\mu}(\epsilon_{1},\epsilon_{2}|\epsilon_{3})\, \in K_{2}\;.
\end{equation}
The expressions for the other channels can be found by relabeling the particles. Note that here $bm_{2}$ equals $b_{2}=[b,m_2]$, since the inputs are annihilated by $b$ due to gauge fixing.

If one uses the explicit expressions for the kinematic maps in the definition of $\mathfrak{n}^{\mu}_{s_{ij}}$, one does not recover the kinematic numerators in the standard form. The numerators written in terms of the currents $\mathfrak{n}^{\mu}_{s_{ij}}$ are related to the numerators in the standard form by momentum conservation. In order to check this, let us use the expressions for the kinematic brackets. First, let us write the $b_{2}$ as
\begin{equation}
    b^{\mu}_{2}(\e_{1},\e_{2})=i\, \e_{12}^{\mu}\,e^{i(p_1+p_2)\cdot x}=i\, \big\{ 2\, \e_{1}\cdot p_{2}\e_{2}^{\mu}+p_{1}^{\mu}\, \e_{1}\cdot \e_{2}-(1 \leftrightarrow 2) \big\}\,e^{i(p_1+p_2)\cdot x}\;.
\end{equation}
Notice that the output of $b_{2}$ belongs to the space of fields. Thus, the kinematic numerator of the $s$-channel in this form is
\begin{equation}
\begin{split}
    n_{s}&=\e_{12\, \mu}\big\{2\e_{3}\cdot (p_{1}+p_{2})\, \e^{\mu}_{4}-2\,  p^{\mu}_{3}\, \e_{3}\cdot \e_{4}+\e^{\mu}_{3}\, \e_{4}\cdot (p_{1}+p_{2}-p_{3})\big\}\\
    &+s\, \big(\e_{1}\cdot\e_{4}\, \e_{2}\cdot \e_{3}-\e_{2}\cdot\e_{4}\, \e_{1}\cdot \e_{3}\big)\\
    &=-\big\{ 2\, \e_{1}\cdot p_{2}\e_{2}^{\mu}+p_{1}^{\mu}\, \e_{1}\cdot \e_{2}-(1 \leftrightarrow 2) \big\}\big\{ 2\, \e_{3}\cdot p_{4}\, \e_{4\, \mu}+p_{3\, \mu}\, \e_{3}\cdot \e_{4}-(3 \leftrightarrow 4) \big\}\\
    &+s\, \big(\e_{1}\cdot\e_{4}\, \e_{2}\cdot \e_{3}-\e_{2}\cdot\e_{4}\, \e_{1}\cdot \e_{3}\big)\;,
\end{split}
\end{equation}
where to arrive at the last equality we used momentum conservation and wrote explicitly the value of $\e_{12}^{\mu}$. The expression in the last equality is the most commonly used for the kinematic numerators. However, in order to investigate their algebraic nature, it turns out to be more convenient to think of them in terms of the currents $\mathfrak{n}_{s_{ij}}^{\mu}$, as we will see later in this section. 

Let us now turn to gravity. We will consider a 4-point tree-level DFT amplitude. In analogy to Yang-Mills theory we impose the gauge and on-shell conditions $b^{+}\Psi=0$  and $B_{1}(\Psi)=0$. We will only take tensors as external particles, which, in combination with the fact that we are dealing with tree-level amplitudes, allows us to discard any possible scalar contributions to the process. For this reason we only consider as external states plane wave solutions given by
\begin{equation}
    e_{\mu\bar\nu}(x,\bar x)=\epsilon_\mu(x)\otimes\bar\epsilon_{\bar\nu}(\bar x)=\varepsilon_{\mu\bar \nu}(p,\bar p)\, e^{i(p\cdot x+\bar p\cdot\bar x)}\,,
\end{equation}
where the polarization tensor $\varepsilon_{\mu\bar\nu}$ of the $i$-th external particle is given by the product of polarization vectors of single-copy elements
\begin{equation}\label{polten}
    \varepsilon_{i\, \mu\bar\nu}(p_{i},\bar p_{i})=\epsilon_{i\, \mu}(p_{i})\,\epsilon_{i\, \bar \nu}(\bar p_{i})\,.
\end{equation}
The two copies of the polarization vectors and momenta obey the gauge fixing and on-shell conditions \eqref{condi}, in addition to the strong constraint $s_{ij}\equiv \bar s_{ij}$.

The 4-point tree-level amplitude for the tensor sector of DFT is given by
\begin{equation}
    \mathcal{M}_{\text{Tree}}^{(4)}=-2\, \kappa^{2}\Big\langle e_{4},[B_{2}(\mathfrak{h}B_{2}( e_{1}, e_{2}), e_{3})+\text{cyclic}]-B_{3}( e_{1}, e_{2}, e_{3})\Big\rangle_{\text{DFT}}\;,
\end{equation}
where $\kappa$ is the gravitational coupling constant, the propagator is $\mathfrak{h}=-\frac{b^{+}}{s_{ij}}$, and the inner product will be defined explicitly below. In order to relate this amplitude to Yang-Mills, it will be helpful to explicitly illustrate how to use the $B_{3}$ in terms of the single-copy maps. To this end, we shall consider as inputs only polarization tensors of the form shown in equation \eqref{polten}. Moreover, we will use the simpler version of $B_{3}$ displayed in equation \eqref{niceB3}. This leads to
\begin{equation}\label{niceB3withvare}
\begin{split}
    B_{3}(e_{1},e_{2},e_{3})=\frac{1}{8}\, b^{-}\Big\{\frac{1}{3}\, &\theta_{3s}\otimes \bar m_{2}\bar m_{2}- \frac{1}{3}\, m_{2}m_{2}\otimes \bar \theta_{3s}+m_{2}b_{2}\otimes \bar m_{3h}+m_{3h}\otimes \bar m_{2}\bar b_{2}\\
    &-m_{3h}\otimes \bar m_{3h}\bar d_{\bar s}\Big\}\, \mathcal{C}\, (\e_{1},\e_{2},\e_{3})\otimes (\bar\e_{1},\bar\e_{2},\bar\e_{3})\;.
\end{split}
\end{equation}
Notice that we have picked a global minus sign coming from the transition
\begin{equation}
    (\e_{1}\otimes\bar \e_{1},\e_{2}\otimes \bar\e_{2},\e_{3}\otimes \bar\e_{3})\to -(\e_{1},\e_{2},\e_{3})\otimes (\bar\e_{1},\bar\e_{2},\bar\e_{3})\;,
\end{equation}
because the elements $\epsilon_i(x)$ have odd degree. Not all the terms in $B_{3}$ contribute to a tree-level scattering amplitude with external tensor particles. To see this, let us take a closer look at the first term in equation \eqref{niceB3withvare}. Omitting the pre-factor we have
\begin{equation}
    b^{-}\theta_{3s}\otimes \bar m_{2}\bar m_{2}\, \mathcal{C}\, (\e_{1},\e_{2},\e_{3})\otimes (\bar\e_{1},\bar\e_{2},\bar\e_{3})=b^{-}\theta_{3s}(\e_{1},\e_{2},\e_{3})\otimes \bar m_{2}(\bar m_{2}(\bar\e_{1},\bar\e_{2}),\bar\e_{3})+\text{cyclic},
\end{equation}
with cyclic denoting the sum over simultaneous cyclic permutations of both barred and un-barred particle labels. Upon looking at the component expressions of the maps involved in the above tensor product (see equations \eqref{m2s} and \eqref{theta2}), it is possible to see that these are scalar quantities that belong to the space of scalar field equations, or currents. For this reason the terms containing $\theta_{3s}$ and $\bar\theta_{3s}$ do not contribute to the amplitude of interest and thus can be ignored. The contributing terms are then 
\begin{equation}
\begin{split}
    -\frac{1}{8}\, b^{-}\Big\{ m_{2}(b_{2}(\e_{1},\e_{2}),\e_{3})\otimes\bar m_{3h}(\bar\e_{1},\bar \e_{2},\bar \e_{3})+m_{3h}(\e_{1},\e_{2},\e_{3})\otimes \bar m_{2}(\bar b_{2}(\bar\e_{1},\bar \e_{2}),\bar\e_{3})\\
    +s\, m_{3h}(\e_{1},\e_{2},\e_{3})\otimes \bar m_{3h}(\bar\e_{1},\bar\e_{2},\bar\e_{3})+\text{cyclic}\Big\}\;,
\end{split}
\end{equation}
where all terms picked a sign due to the three polarization vectors passing through maps of odd degree ($b_{2},\bar b_{2},m_{3h},\bar m_{3h}$), and the last term picked an extra minus sign coming from the fact that in momentum space $d_{s_{ij}}$ translates into $-s_{ij}$. Thus, the 4-point amplitude can be written in terms of the kinematic maps of Yang-Mills as
\begin{equation}
\begin{split}
         \mathcal{M}_{\text{Tree}}^{(4)}=-\frac{\kappa^{2}}{4}\, \Big\langle \e_{4}\otimes\bar \e_{4},\frac{1}{s}\, b^{-}&\left[ m_{2}(b_{2}(\e_{1},\e_{2}),\e_{3})\otimes \bar m_{2}(\bar b_{2}(\bar \e_{1},\bar \e_{2}),\bar\e_{3})\right.\\
         &\left. +s\, m_{2}(b_{2}(\e_{1},\e_{2}),\e_{3})\otimes \bar m_{3h}(\bar \e_{1},\bar\e_{2},\bar \e_{3})\right.\\
         &\left. +s\, m_{3h}(\e_{1},\e_{2},\e_{3})\otimes \bar m_{2}(\bar b_{2}(\bar\e_{1},\bar\e_{2}),\bar\e_{3})\right.\\
         &\left. +s^{2}\, m_{3h}(\e_{1},\e_{2},\e_{3})\otimes \bar m_{3h}(\bar\e_{1},\bar \e_{2},\bar\e_{3}) \right]+\text{cyclic} \Big\rangle_{\text{DFT}}\;,
\end{split}
\end{equation}
where for the exchange contribution we used 
\begin{equation}
    B_{2}b^{+}B_{2}=-\frac{1}{8}\, b^{-}m_{2}b_{2}\otimes \bar m_{2}\bar b_{2}\;,
\end{equation}
and the strong constraint $s_{ij}\equiv \bar s_{ij}$. 

Let us now make contact with the BCJ double copy of amplitudes. To this end, we examine the factorization property of the DFT inner product, which, for tensors, is defined as
\begin{equation}
    \big\langle \varepsilon_{i},\mathcal{J}_{j} \big\rangle_{\text{DFT}}=\varepsilon_{i\, \mu\bar\nu}(p_{i},\bar p_{i})\, \mathcal{J}^{\mu\bar\nu}_{j}(p_{j},\bar p_{j})\;,
\end{equation}
where we omitted the double momentum-conserving delta function. The currents belong to the space of field equations of DFT, and hence can be expressed as $\mathcal{J}^{\mu\bar\nu}_{i}=-b^{-}\, J^{\mu}_{i}\otimes \bar J^{\bar\nu}_{i}$. For this reason, it is possible to factorize the DFT inner product as
\begin{equation}
   -\big\langle \e_{i}\otimes \bar\e_{i},b^{-}\, J_{j}\otimes \bar J_{j} \big\rangle_{\text{DFT}}=\big\langle \e_{i},J_{j}\big\rangle_{\text{YM}}\,  \big\langle\bar \e_{i},\bar J_{j}\big\rangle_{\text{YM}}\,.
\end{equation}
Using this relation between the inner products, it is  possible to see that the DFT amplitude can be written as
\begin{equation}
\begin{split}
    \mathcal{M}_{\text{Tree}}^{(4)}&=\frac{\kappa^{2}}{4}\Bigg\{ \frac{\big\langle\e_{4},\mathfrak{n}_{s}\big\rangle_{\text{YM}}\,  \big\langle \bar \e_{4},\bar {\mathfrak{n}}_{s} \big\rangle_{\text{YM}}}{s} \Bigg\}+\text{cyclic}\\
    &=\frac{\kappa^2}{4}\left\{ \frac{n_{s}\, \bar n_{s}}{s}+\frac{n_{t}\, \bar n_{t}}{t}+\frac{n_{u}\, \bar n_{u}}{u} \right\}.
\end{split}
\end{equation}
This amplitude agrees with the expectation from the BCJ double copy because, as it is straightforward to notice, this amplitude can be obtained by exchanging color and kinematics \`a la BCJ, namely exchanging $c_{s_{ij}}\to \bar n_{s_{ij}}$ and $g_{\rm YM}\to \frac{\kappa}{2}$ in the Yang-Mills amplitude \eqref{forbidden}. Moreover, if we solve the strong constraint by setting $p_{\mu}=\bar p_{\bar\mu}$, one recovers the 4-point amplitude of $N=0$ supergravity. 

The BCJ double copy requires the kinematic numerators to obey the so-called \textit{kinematic Jacobi identity}. This relation guarantees gauge invariance of the gravity amplitude, and thus ensures its consistency. Now we argue that this relation follows in a straightforward manner from the homotopy Poisson relation. Let us recall the homotopy Poisson relation in an input free form:
\begin{equation}
    [b,m_{2}m_{2}]-3m_{2}b_{2}\pi-[m_{1},\theta_{3}]-m_{3h}(d_{\square}-3\, d_{s}\, \pi)=0\;.
\end{equation}
Since we want to relate this equation to the kinematic numerators, all the inputs that we will consider are polarization vectors obeying the gauge and on-shell conditions. Additionally, in order to recover the kinematic numerators from this equation, it is necessary to take the inner product of the Poisson relation with a polarization vector $\e_{4\, \mu}$. Doing so the first term vanishes because all the polarization vectors are annihilated by $b$. The third and fourth term vanish because all the the polarization vectors are on-shell, and hence are annihilated by $m_{1}$ and $\square$. Notice that the second term in  combination with the last term is the cyclic sum of the currents $\mathfrak{n}_{s_{ij}}^{\mu}$. Thus, upon taking the inner product with a polarization vector $\e_{4}$ we obtain
\begin{equation}
    n_{s}+n_{t}+n_{u}=0\;.
\end{equation}

\subsection{Three-bracket of the gauge algebra} 
As an additional and independent concrete example, let us examine the gauge algebra of DFT. In the following it is assumed that no gauge fixing condition is imposed. In \cite{Bonezzi:2022yuh} we found the 2-bracket of two DFT gauge parameters and learned how the kinematic structure of Yang-Mills is a fundamental building block of the gauge algebra of DFT. The 2-bracket between two gauge parameters is given by
\begin{equation}
    B_{2}(\Lambda_{1},\Lambda_{2})=\begin{pmatrix}
    \lambda^{\mu}_{12}\\
    \bar \lambda^{\bar\mu}_{12}\\
    \eta_{12}
    \end{pmatrix}\;,
\end{equation}
where the components are 
\begin{equation}\label{compB2}
\begin{split}
    \lambda_{12}^{\mu}&=-\frac{1}{4}(\lambda_{1}\bullet\lambda_{2})^{\mu}-\frac{1}{4}\bar\del_{\bar\nu}(\lambda_{1}^{\mu}\, \bar\lambda_{2}^{\bar\nu})+\frac{1}{4}\bar\del_{\bar\nu}(\lambda_{2}^{\mu}\, \bar\lambda_{1}^{\bar\nu})\;,\\   
    \bar\lambda_{12}^{\bar\mu}&=\frac{1}{4}(\bar\lambda_{1}\bullet\bar\lambda_{2})^{\bar\mu}+\frac{1}{4}\del_{\nu}(\bar\lambda_{1}^{\bar \mu}\, \lambda_{2}^{\nu})-\frac{1}{4}\del_{\nu}(\bar\lambda_{2}^{\bar\mu}\, \lambda_{1}^{\nu})\;,\\
    \eta_{12}&=-\frac{1}{2}\del_{\mu}\bar\del_{\bar\nu}\big(\lambda_{1}^{\mu}\, \bar\lambda_{2}^{\bar\nu}- \lambda_{2}^{\mu}\, \bar\lambda_{1}^{\bar\nu}\big)\;,
\end{split}
\end{equation}
where the product $\bullet$ is defined as
\begin{equation}\label{YMbullet}
  (v\bullet w)_{\mu} = v^{\nu}\partial_{\nu}w_{\mu} + (\partial_{\mu}v^{\nu}-\partial^{\nu}v_{\mu})w_{\nu}
  +(\partial_{\nu}v^{\nu})w_{\mu}-(v \leftrightarrow w)\;.
\end{equation}
 
Let us now look at the 3-bracket. It is worth mentioning that the general 3-bracket derived earlier cannot take as an input an $\eta$ with any other two gauge parameters. The reason is that $\eta$ can be written as $\eta=b^{-}(\varphi\otimes\bar\varphi)$. Since the auxiliary field $\varphi$ cannot be the input of $m_{2}$ nor $m_{3}$, $\eta$ can only be taken as an input in the terms of $B_{3}$ that have either a $\theta_{3s}$ or a $\bar\theta_{3s}$. The two vector gauge parameters can be expressed in terms of Yang-Mills elements as $\lambda_{\mu}=A_{\mu}\otimes\bar \lambda $ and $\bar \lambda_{\bar\mu}=\lambda\otimes \bar A_{\bar\mu}$. Hence, the only possibility of $\eta$ being an input of $B_{3}$ with two other gauge parameters is if there exists a $\theta_{3s}$ that takes as inputs $\varphi$ with either $\lambda$ or $A_{\mu}$. However, no such $\theta_{3}$ exists (see appendix \ref{AppYM}). This rules out the appearance of $\eta$ in the 3-bracket of three gauge parameters and thus we can only take the vector components as inputs.

Consider now one gauge parameter $\lambda_{\mu}$ and two $\bar \lambda_{\bar\mu}$ as inputs. Such an arrangement of inputs is also impossible, because the $\theta_{3s}$ and thus $\bar\theta_{3s}$ with two fields and one gauge parameter, and one field and two gauge parameters vanish. Similar arguments follow for two $\lambda_{\mu}$ and one $\bar\lambda_{\bar\mu}$. For this reason, the only viable option is to consider as inputs three vector parameters of the same chirality. Notice that with three $\lambda_{\mu}$ as inputs the only non-vanishing contributions are the terms containing $\theta_{3s}$ and $\bar\theta_{3s}$. This follows because $m_{3}$ with three Yang-Mills gauge parameters vanishes, and $b_{2}$ with two Yang-Mills gauge parameters is zero because of degree reasons. Thus, explicitly, the $B_{3}$ acting on three DFT parameters is given by
\begin{equation}\label{B3lam}
\begin{split}
    B_{3}(\Lambda_{1},\Lambda_{2},\Lambda_{3})=-\frac{1}{8}\, b^{-}&\big\{ \theta_{3s}(A_{1},A_{2},A_{3})\otimes \bar m_{2}\bar m_{2}\bar\pi (\bar \lambda_{1},\bar\lambda_{2},\bar \lambda_{3})\big\}\\
    +\frac{1}{8}\, b^{-}&\big\{ m_{2}m_{2}\pi (\lambda_{1},\lambda_{2},\lambda_{3})\otimes \bar\theta_{3s}(\bar A_{1},\bar A_{2},\bar A_{3}) \big\}\;.
\end{split}
\end{equation}
The two single-copy maps are in components
\begin{equation}
\begin{split}
    \theta_{3s}(A_{1},A_{2},A_{3})&\stackrel{[123]}{=}6\, A_{1}\cdot\del A_{2}\cdot A_{3}\, \in K_{1}\;,\\
    \bar m_{2}\bar m_{2}\bar\pi(\bar\lambda_{1},\bar\lambda_{2},\bar\lambda_{3})&=\bar\lambda_{1}\, \bar\lambda_{2}\, \bar\lambda_{3}\, \in \bar K_{0}\;.
\end{split}
\end{equation}
where the $[123]$ signifies implicit antisymmetrization over the labels with strength one. The action of $b^{-}$ in the definition of $B_{3}$ is determined by the action of $b$ and $\bar b$ on the single-copy maps. They act as
\begin{equation}
\begin{split}
    b\,\theta_{3s}(A_{1},A_{2},A_{3})&\stackrel{[123]}{=}6\, A_{1}\cdot\del A_{2}\cdot A_{3}\, \in K_{0}\;,\\
    \bar b\,\bar m_{2}\bar m_{2}\bar\pi (\bar\lambda_{1},\bar\lambda_{2},\bar\lambda_{3})&=0\;.
\end{split}
\end{equation}
Following the same arguments for the other term one obtains
\begin{equation}\label{3gauge}
\begin{split}
    B_{3}(\Lambda_{1},\Lambda_{2},\Lambda_{3})&\stackrel{[123]}{=}-\frac{3}{8}\big\{ A_{1}\cdot\del A_{2}\cdot A_{3}\otimes \bar\lambda_{1}\, \bar\lambda_{2}\, \bar\lambda_{3}+\lambda_{1}\, \lambda_{2}\, \lambda_{3}\otimes \bar A_{1}\cdot \bar\del \bar A_{2}\cdot\bar A_{3}  \big\}\\
    &\stackrel{[123]}{=}-\frac{3}{8}\big\{ (A_{1\, \mu}\otimes \bar\lambda_{1})\del^{\mu}(A^{\rho}_{2}\otimes \bar \lambda_{2})(A_{3\, \rho}\otimes \bar\lambda_{3})+(\lambda_{1}\otimes \bar A_{1\, \bar\mu})\bar\del^{\bar\mu}(\lambda_{2}\otimes \bar A^{\bar\rho}_{2})(\lambda_{3}\otimes \bar A_{3\, \bar\rho})\big\}\\
    &\stackrel{[123]}{=}-\frac{3}{8}\big\{\lambda_{1}\cdot\del\lambda_{2}\cdot\lambda_{3}+\bar\lambda_{1}\cdot\bar\del\bar\lambda_{2}\cdot\bar\lambda_{3}\big\}\,\in K_{0}\otimes \bar K_{0}\;,
\end{split}
\end{equation}
which in $\cV_{\rm strong}$ is the space of gauge-for-gauge parameters.
The gauge algebra of double field theory is an $L_{\infty}$ algebra, and hence the 2-brackets should satisfy the homotopy Jacobi relation
\begin{equation}
    B_{2}(B_{2}(\Lambda_{1},\Lambda_{2}),\Lambda_{3})+B_{2}(B_{2}(\Lambda_{2},\Lambda_{3}),\Lambda_{1})+B_{2}(B_{2}(\Lambda_{3},\Lambda_{1}),\Lambda_{2})+[B_{1},B_{3}](\Lambda_{1},\Lambda_{2},\Lambda_{3})=0.
\end{equation}
In order to check the consistency of our result for the 3-bracket with three gauge parameters, we verified the above identity by using the component form of the 2-brackets \eqref{compB2} and 3-bracket \eqref{3gauge}. In its standard formulation, the gauge algebra of DFT is a particular type of $L_{\infty}$ algebra associated to the Courant algebroid \cite{Roytenberg:1998vn,Hull:2009zb}. The 3-bracket derived in this paper is not the same found in the standard formulation of DFT. However, in \cite{Bonezzi:2022yuh} we found that the gauge transformations obtained by using the 2-bracket \eqref{B222} agree with the ones of \cite{Hull:2009mi} upon a field-dependent parameter redefinition. This guarantees that the 3-bracket found here from \eqref{niceB3} is equivalent to the standard one up to an $L_\infty$ morphism.

\section{Conclusions and Outlook}

In this paper we have generalized a recent off-shell double copy construction of gravity (in the form of 
double field theory) from Yang-Mills theory \cite{Diaz-Jaramillo:2021wtl,Bonezzi:2022yuh} by giving 
a gauge invariant and local prescription up to and including quartic order. 
To this end we used the homotopy algebra formulation of gauge field theories, starting 
from the $L_{\infty}$ algebra of  Yang-Mills theory in a particular formulation inspired by 
string field theory, and stripping off color in order to arrive at a $C_{\infty}$ algebra on the 
Yang-Mills kinematic vector space ${\cal K}$. This structure was used recently in order to define gravity via  double copy 
to cubic order, but as reported here the transition to quartic order 
requires a much larger algebra to be present just in Yang-Mills theory proper. 
We find that an algebra proposed by Reiterer in \cite{Reiterer:2019dys}, and called BV$_{\infty}^{\square}$,  
is also realized  in our formulation of Yang-Mills theory. More precisely, we prove this up to and including 
trilinear maps, as needed to quartic order in field theory, and we compute the corresponding 3-brackets 
of double field theory purely from these algebraic structures of Yang-Mills theory.

The most important outstanding problem, and the missing step toward  a  construction 
of full-fledged  gravity from Yang-Mills theory, is to display the BV$_{\infty}^{\square}$ algebra on ${\cal K}$, 
and the associated $L_{\infty}$ algebra on ${\cal K}\otimes \bar{\cal K}$, to all orders. 
In practice this step will require a much deeper understanding of why these structures are 
present in Yang-Mills theory, which to this order we have verified by explicit brute-force computations. 
It would therefore be highly desirable to arrive at some sort of `derived' construction, where this algebra is 
obtained from something much simpler. It is striking that an algebra as complex 
as gravity itself appears to be present already in pure Yang-Mills theory. 
Perhaps we can learn a lot more about classical and quantum gravity by just studying Yang-Mills theory 
more closely.

\subsection*{Acknowledgements}

We would like to thank Maor Ben-Shahar, Tomas Codina, Lucia M. Garozzo, Henrik Johansson, Ricardo Monteiro, Nathan Moynihan, Allison Pinto, Jan Plefka and Oliver Schlotterer for useful discussions.

This work is funded   by the European Research Council (ERC) under the European Union's Horizon 2020 research and innovation programme (grant agreement No 771862)
and by the Deutsche Forschungsgemeinschaft (DFG, German Research Foundation), ``Rethinking Quantum Field Theory", Projektnummer 417533893/GRK2575.   

\appendix
\section{Explicit maps of Yang-Mills theory}\label{AppYM}

In this appendix we collect the relevant maps of the ${\rm BV}_\infty^\B$ algebra of Yang-Mills. We start by recalling the components of the graded vector space $\cK$:
\begin{equation}
\begin{tikzcd}[row sep=2mm]
&K_{0}\arrow{r}{m_{1}} & K_{1}\arrow{r}{m_{1}} & K_2\arrow{r}{m_{1}} & K_3\\
\cK^{(0)}:&\lambda& A_\mu & E\\
\cK^{(1)}: & &\arrow{ul}{b}\varphi&\arrow{ul}{b}E_\mu &\arrow{ul}{b}\cN
\end{tikzcd}\;,
\end{equation}
together with the action of the differential $m_{1}$:
\begin{equation}
m_{1}(\lambda)=\bpm\del_\mu\lambda\\\square\lambda\epm\in K_{1}\;,\quad m_{1}\bpm A_\mu\\\varphi\epm=\bpm\del\cdot A-\varphi\\\square A_\mu-\del_\mu\varphi\epm\in K_2\;,\quad m_{1}\bpm E\\ E_\mu\epm=\square E-\del^\mu E_\mu\in K_3\;,  
\end{equation}
and of the $b$ operator:
\begin{equation}
b\bpm A_\mu\\\varphi\epm=\varphi\in K_{0}\;,\quad b\bpm E\\E_\mu\epm=\bpm E_\mu\\0\epm\in K_{1}\;,\quad b\cN=\bpm\cN\\0\epm\in K_2\;,    
\end{equation}
from which one can easily verify that $m_{1}\,b+b\,m_{1}=\B$.
We now recall from \cite{Bonezzi:2022yuh} the explicit form of the non-vanishing $C_\infty$ 2-products $m_2$:
\begin{equation}\label{m2s}
\begin{split}
m_2\big(\lambda_1, \lambda_2\big)&=\lambda_1 \lambda_2\in K_0\;,\quad\hspace{20mm} m_2\big(\cA, \lambda\big)=\bpm A^\mu\lambda\\[2mm]\del_\nu(A^\nu \lambda)\epm\in K_1 \;,\\[3mm] 
m_2\big(\cA_1, \cA_2\big)&=\bpm0\\[2mm]m_2^\mu(A_1, A_2)\epm\in K_2\;,\quad m_2\big(\lambda, \cE\big)=\bpm0\\[2mm]\lambda( E^\mu-\del^\mu E)\epm\in K_2\;,\\[3mm] 
m_2\big(\cA, \cE\big)&=A_\mu(\del^\mu E-E^\mu)\in K_3\;,\quad\hspace{3mm} m_2\big(\lambda, \cN\big)=\lambda\, \cN\in K_3\;,
\end{split}    
\end{equation}
where $m_2^\mu(A_1, A_2)$, encoding the color-stripped cubic vertex, is given by 
\begin{equation}
m_2^\mu(A_1, A_2)\stackrel{[12]}{=}2\,\del\cdot A_1 A_2^\mu+4\,A_1\cdot\del A_2^\mu+2\,\del^\mu A_1\cdot A_2 \;,   
\end{equation}
with $[12]$ denoting implicit antisymmetrization with strength one.
The only non-vanishing three product $m_3$ is between three fields:
\begin{equation}
\begin{split}
m_3\big(\cA_1, \cA_2, \cA_3\big)&=\bpm0\\[2mm] m^\mu_3\big(A_1, A_2,A_3\big)\epm\in K_2 \;,\\
m^\mu_3\big(A_1, A_2,A_3\big)&=A_1\cdot A_2\,A_3^\mu+A_3\cdot A_2\,A_1^\mu-2\,A_1\cdot A_3\,A_2^\mu \;,   
\end{split}
\end{equation}
and corresponds to the color-stripped quartic vertex.
We do not give the explicit form of the 2-brackets $b_2$, since they can be straightforwardly computed by taking a $b-$commutator of $m_2$:
\begin{equation}
b_2(u_1,u_2)=b\,m_2(u_1,u_2)-m_2(b u_1,u_2)-(-)^{u_1u_2}m_2(u_1,bu_2)\;.    
\end{equation}
By direct computation of the failure of the compatibility condition:
\begin{equation}
\begin{split}
k_3(u_1,u_2|u_3)&=b_2\big(m_2(u_1,u_2),u_3\big)-(-)^{u_1(u_2+u_3)}m_2\big(b_2(u_2,u_3),u_1\big)\\
&\hspace{35mm}-(-)^{u_3(u_1+u_2)}m_2\big(b_2(u_3,u_1),u_2\big)\;,    
\end{split}
\end{equation}
we proved the deformed homotopy Poisson relation \eqref{homPoiss}. Here we give all the non-vanishing Poisson homotopy maps $\theta_3$. The following ones are purely graded symmetric:
\begin{equation}\label{theta1}
\begin{split}
\theta_{3}(\cE,\lambda_1,\lambda_2)&=\lambda_1\lambda_2\,E\in K_0\;,\\
\theta_{3}(\cN,\lambda_1,\lambda_2)&=-\bpm0\\\lambda_1\lambda_2\,\cN\epm\in K_1\;,\\
\theta_{3}(\cE,\cA,\lambda)&=\bpm \,A_\mu\lambda E\\A^\nu E_\nu\lambda+A^\nu\del_\nu\lambda E+\varphi\lambda E\epm\in K_1\;,\\
\theta_{3}(\lambda,\cE_1,\cE_2)&=\bpm\lambda E_1E_2\\\lambda (E_1^\mu E_2+E_2^\mu E_1)\epm\in K_2\;,\\
\theta_{3}(\lambda,\cE,\cN)&=\lambda E\cN\in K_3\;,\\
\theta_{3}(\cA,\cE_1,\cE_2)&=\varphi E_1E_2+2\,A^\mu\del_\mu (E_1 E_2)-A_\mu (E_1^\mu E_2+E_2^\mu E_1)\in K_3\;.
\end{split}    
\end{equation}
The last two $\theta_3$ maps have both a graded symmetric and a hook part, which we give explicitly:
\begin{equation}\label{theta2}
\begin{split}
\theta_3(\cA_1,\cA_2|\cA_3)&=\theta_{3s}(\cA_1,\cA_2,\cA_3)+\theta_{3h}(\cA_1,\cA_2|\cA_3)\in K_1\;,\quad{\rm where}\\
\theta_{3s}(\cA_1,\cA_2,\cA_3)&\stackrel{[123]}{=}6\,\bpm0\\A_1\cdot\del A_2\cdot A_{3}\epm\;,\quad
\theta_{3h}(\cA_1,\cA_2|\cA_3)\stackrel{[12]}{=}-2\,\bpm A_1^\mu A_2\cdot A_3\\0\epm\;,
\end{split}    
\end{equation}
and
\begin{equation}\label{theta3} 
\begin{split}
\theta_3(\cA_1,\cA_2|\cE)&=\theta_{3s}(\cA_1,\cA_2,\cE)+\theta_{3h}(\cA_1,\cA_2|\cE)\in K_2\;,\\
\theta_3(\cE,\cA_1|\cA_2)&=\theta_{3s}(\cA_1,\cA_2,\cE)+\theta_{3h}(\cE,\cA_1|\cA_2)\in K_2\;,\quad{\rm where}\\[3mm]
\theta_{3s}(\cE,\cA_1,\cA_2)&\stackrel{[12]}{=}2\,\bpm0\\A^\mu_2A^\nu_1\big(E_\nu+\del_\nu E\big)+E\,\big(2\,A_1\cdot\del A_2^\mu+\del^\mu A_1\cdot A_2\big)+A^\mu_2\varphi_1 E\epm\;,\\[3mm]
\theta_{3h}(\cA_1,\cA_2|\cE)&\stackrel{[12]}{=}-2\,\bpm0\\A^\mu_1A_2\cdot E\epm\;,\quad \theta_{3h}(\cE,\cA_1|\cA_2)=\bpm0\\A^\mu_1A_2\cdot E-E^\mu A_1\cdot A_2\epm\;.
\end{split}    
\end{equation}
We do not give the 3-brackets $b_3$, since they can be computed directly from $b_3=-[b,\theta_{3s}]$.

\section{From ${\rm BV}$ to ${\rm BV}_\infty$ algebras}\label{AppBV}

\subsection*{dgBV Algebra}

Here we will review the structure of ${\rm BV}_\infty$ up to three inputs. For an all order description, see \cite{Galvez-Carrillo:2009kic}. As with other types of homotopy algebras, this structure arises naturally when introducing a differential compatible with the BV structure. In the paper, we encounter a further generalization of ${\rm BV}_\infty$ algebras, called ${\rm BV}^\square_\infty$ algebras.

A BV algebra has an associative and graded commutative product $m_2$ together with an operator $b$ that is second order with respect to that product. The operator $b$ induces a Lie bracket $b_2$ defined by
\begin{equation}
b_2(u_1,u_2) = (-)^{u_1}[b,m_2](u_1,u_2) := (-)^{u_1} bm_2(u_1,u_2) - (-)^{u_1} m_2(bu_1,u_2) - m_2(u_1,bu_2) \; .
\end{equation}
Here we used the sign convention commonly used in BV algebras. For the generalization used in the main text, it will be more convenient to change the sign convention. We make the redefinition $b_2(u_1,u_2) \mapsto (-)^{u_1}b_2(u_1,u_2)$, so that $b_2$ becomes graded symmetric, i.e.
\begin{equation}\label{BVsym}
b_2(u_1,u_2) = (-)^{u_1 u_2}b_2(u_2,u_1) \; .
\end{equation}
Since $b$ is second order, we have that $b_2(u_1,-)$ is a derivation of $m_2$ from the left. This means that
\begin{equation}\label{bsecondorder}
b_2(u_1,m_2(u_2,u_3)) = m_2(b_2(u_1,u_2),u_3) + (-)^{u_2(u_1+1)}m_2(u_2,b_2(u_1,u_3)) \; .
\end{equation}
We note that this is equivalent to
\begin{equation}
b_2(m_2(u_1,u_2),u_3) = (-)^{u_1}m_2(u_1,b_2(u_2,u_3)) + (-)^{u_2u_3}m_2(b_2(u_1,u_3),u_2) \; .
\end{equation}
From $b^2 = 0$, it then further follows that
\begin{equation}
[b,b_2](u_1,u_2) := b b_2(u_1,u_2) + (-)^{u_1} b_2(bu_1,u_2) + (-)^{u_1}b_2(u_1,bu_2)  = 0 \; ,
\end{equation}
i.e., $b$ is a derivation of $b_2$. We can use this on \eqref{bsecondorder} to find that
\begin{equation}\label{BVJac}
b_2(b_2(u_1,u_2),u_3) +  (-)^{u_2u_3}b_2(b_2(u_1,u_3),u_2) + (-)^{u_1(u_2+u_3)} b_2(b_2(u_2,u_3),u_1) = 0 \; .
\end{equation}

Equations \eqref{BVsym} and \eqref{BVJac} tell us that $b_2$ defines a graded Lie algebra. Toghether with $m_2$ we have a Gerstenhaber algebra. The condition that $b$ is second order is a graded version of the Poisson relation. However, a BV algebra has more structure than a Gerstenhaber algebra, because in the BV case, the bracket $b_2$ comes from a second order operator, in our case $b$.

A BV algebra becomes a dgBV algebra, once we introduce another differential $m_{1}$ and demand that it commutes with both $b$ and $m_2$ (i.e. it is a derivation with respect to $m_2$). It follows that it is also a derivation of $b_2$. In this sense, a dgBV algebra combines the concept of a dgLie algebra with that of a dg commutative algebra. $(m_{1},b_2)$ form a dgLie algebra, while $(m_{1},m_2)$ form a dg commutative algebra. The compatibility of $m_2$ with $b_2$ then relates these two structures.

\subsection*{Homotopy BV Algebra}
\subsubsection*{Associative Product}

We will now relax the condition that $b$ is of second order. As with other homotopy algebras like $L_\infty$ and $C_\infty$ algebras, we do not want to give up this condition completely. We want it to hold \emph{up to homotopy}. This means that the conditions \eqref{bsecondorder} are relaxed to
\begin{equation}\label{2ndorderuptoh}
[m_{1},\theta_3](u_1,u_2,u_3) = b_2(m_2(u_1,u_2),u_3) - (-)^{u_1}m_2(u_1,b_2(u_2,u_3)) - (-)^{u_2u_3}m_2(b_2(u_1,u_3),u_2) \; .
\end{equation}
We have that $[m_{1},\theta_3]$ is graded symmetric in the first two entries, i.e.
\begin{equation}
[m_{1},\theta_3](u_1,u_2,u_3) = (-)^{u_1u_2}[m_{1},\theta_3](u_2,u_1,u_3) \; .
\end{equation}
Using associativity of $m_2$, it also follows that it is graded symmetric in the last two entries, i.e.
\begin{equation}
[m_{1},\theta_3](u_1,u_2,u_3) = (-)^{u_2u_3}[m_{1},\theta_3](u_1,u_3,u_2) \; .
\end{equation}
We demand that $\theta_3$ has the same symmetry properties.

Recall that we used the $b$-commutator $[b,-]$ to prove that $b_2$ satisfies the graded Jacobi identity. The same can be done here, but now we find
\begin{equation}\label{inducedb3}
\begin{split}
0 = &[m_{1},[b,\theta_3]](u_1,u_2|u_3) - b_2(b_2(u_1,u_2),u_3) \\
& - (-)^{u_1(u_2+u_3)} b_2(b_2(u_2,u_3),u_1) - (-)^{u_2 u_3}b_2(b_2(u_1,u_3),u_2) \; .
\end{split}
\end{equation}
This relation tells us that $b_2$ satisfies the Jacobi identity up to the homotopy $b_3(u_1,u_2,u_3) := -[b,\theta_3](u_1,u_2,u_3)$. Therefore, assuming second order of $b$ only up to homotopy, the algebra naturally contains an $L_\infty$ algebra, at least up to the three brackets. Note that we could also say that $m_2$ and $b_2$ satisfy the Poisson identity up to homotopy.

\subsubsection*{Associativity up to Homotopy}

We just saw that relaxing the second order condition on $b$ naturally leads to an $L_\infty$ algebra, instead of merely a dgLie algebra. This suggests that we could also relax the associativity condition on $m_2$, so that it only holds up to homotopy. Explicitly,
\begin{equation}\label{ComuptpHom}
 m_2(m_2(u_1,u_2),u_3) - m_2(u_1,m_2(u_2,u_3))  = [m_{1},m_3](u_1,u_2,u_3) \; .
\end{equation}
On the right hand side, we introduced a three product $m_3$. We need to demand that it vanishes on signed shuffles, by which we mean that
\begin{equation}
m_3(u_1,u_2,u_3) - (-)^{u_1 u_2}m_3(u_2,u_1,u_3) + (-)^{u_1(u_2 + u_3)}m_3(u_2,u_3,u_1) = 0 \; .
\end{equation}
This is necessary, since the left hand side of \eqref{ComuptpHom} satisfies this identity.

As we noted before, in order for $\theta_3$ to be symmetric in the last two entries, associativity of $m_2$ is necessary. Since this symmetry is no longer guaranteed, we now write $\theta_3(u_1,u_2|u_3)$ instead of $\theta_3(u_1,u_2,u_3)$. The failure for it to be symmetric in $u_2$ and $u_3$ is given by
\begin{equation}
\begin{split}
&[m_{1},\theta_3](u_1,u_2|u_3) - (-)^{u_2u_3}[m_{1},\theta_3](u_1,u_3|u_2)  \\
&= [b,m_2(m_2\otimes 1)](u_1,u_2,u_3) - (-)^{u_2u_3 + u_1 + u_2}[b,m_2(m_2\otimes 1)](u_1,u_3,u_2) \; \\
&= (-)^{u_1u_2} [b,[m_{1},m_3]](u_2,u_1,u_3) = -(-)^{u_1u_2}[m_{1},[b,m_3]](u_2,u_1,u_3) \; .
\end{split}
\end{equation}
In the last step we used that $m_{1}$ commutes with $b$. Note that both sides are $m_{1}$-exact. Therefore, it makes sense to demand that
\begin{equation}
[b,m_3](u_1,u_2,u_3) = -\theta_3(u_1,u_2|u_3)  + (-)^{u_1(u_2+u_3) + u_1 + u_3}\theta_3(u_2,u_3|u_1) \; .
\end{equation}
As a crosscheck, one can show that the right hand side vanishes on signed shuffles, so this is consistent with the left hand side containing $m_3$. This identity connects the $m_3$ of the $C_\infty$ structure to the homotopy of the Poisson identity.

We just saw that the non-symmetric part of $\theta_3$ is a $b$-commutator. Therefore, when defining $b_3(u_1,u_2,u_3) = -[b,\theta_3](u_1,u_2|u_3)$, this part drops out. We can still define $b_3$ as in \eqref{inducedb3} and find that it has the correct symmetry properties.

\providecommand{\href}[2]{#2}\begingroup\raggedright\endgroup

\end{document}